\documentclass[12pt]{article}

\usepackage{newtxtext,newtxmath}

\usepackage{graphicx}

\usepackage[letterpaper,margin=1in]{geometry}

\renewenvironment{abstract}
	{\quotation}
	{\endquotation}

\date{}

\makeatletter
\renewcommand{\fnum@figure}{\textbf{Figure \thefigure}}
\renewcommand{\fnum@table}{\textbf{Table \thetable}}
\makeatother

\usepackage{scicite}

\usepackage{url}

\newcommand{\ef}{$E_{\rm F}$}

\def\scititle{
	Observation of mirror-odd and mirror-even spin texture in ultrathin epitaxially strained RuO$_2$ films
}
\title{\bfseries \boldmath \scititle}

\author{
	Yichen Zhang$^{1\dagger}$,
	Seung Gyo Jeong$^{2\dagger}$,
        Luca Buiarelli$^{2}$,
        Seungjun Lee$^{3,4}$,\and
        Yucheng Guo$^{1}$,
        Jiaqin Wen$^{1}$,
        Hang Li$^{5}$,
        Sreejith Nair$^{2}$,
        In Hyeok Choi$^{6}$,\and
        Zheng Ren$^{1}$,
        Ziqin Yue$^{1,7}$,
        Jounghoon Hyun$^{1}$,
        Tieqiong Zhang$^{1,7}$,\and
        Alexei Fedorov$^{8}$,
        Sung-Kwan Mo$^{8}$,
        Hojoon Lim$^{9,10}$,
        Adrian Hunt$^{9}$,\and
        Iradwikanari Waluyo$^{9}$,
        Junichiro Kono$^{1,11,12,13}$,
        Ján Minár$^{14}$,\and
        Jong Seok Lee$^{6}$,
        Tony Low$^{3}$,
        Turan Birol$^{2}$,
        Rafael M. Fernandes$^{15,16}$,\and
        Milan Radovic$^{5\ast}$,
        Bharat Jalan$^{2\ast}$,
	Ming Yi$^{1,13,17\ast}$\and
	\small$^{1}$Department of Physics and Astronomy, Rice University, Houston, Texas 77005, USA\and
	\small$^{2}$Department of Chemical Engineering and Materials Science, University of Minnesota-Twin Cities,\and\small Minneapolis, Minnesota 55455, USA\and
        \small$^{3}$Department of Electrical and Computer Engineering, University of Minnesota-Twin Cities, Minneapolis,\and\small Minnesota 55455, USA\and
        \small$^{4}$Department of Applied Physics, Kyung Hee University, Yongin 17104, Republic of Korea\and
        \small$^{5}$Photon Science Division, Paul Scherrer Institute, Villigen 5232, Switzerland\and
        \small$^{6}$Department of Physics and Photon Science, Gwangju Institute of Science and Technology (GIST),\and\small Gwangju 61005, Republic of Korea\and
        \small$^{7}$Applied Physics Graduate Program, Smalley-Curl Institute, Rice University, Houston, Texas 77005, USA\and
        \small$^{8}$Advanced Light Source, Lawrence Berkeley National Laboratory, Berkeley, California 94720, USA\and
        \small$^{9}$National Synchrotron Light Source II, Brookhaven National Laboratory, Upton, New York 11973, USA\and
        \small$^{10}$Department of Integrative Energy, Myongji University, Yongin 17058, Republic of Korea\and
        \small$^{11}$Department of Electrical and Computer Engineering, Rice University, Houston, Texas 77005, USA\and
        \small$^{12}$Department of Materials Science and NanoEngineering, Rice University, Houston, Texas 77005, USA\and
        \small$^{13}$Smalley-Curl Institute, Rice University, Houston, Texas 77005, USA\and
        \small$^{14}$New Technologies Research Center, University of West Bohemia, Plzen, 301 00, Czech Republic\and
        \small$^{15}$Department of Physics, The Grainger College of Engineering, University of Illinois Urbana-Champaign,\and\small Urbana, Illinois 61801, USA\and
        \small$^{16}$Anthony J. Leggett Institute for Condensed Matter Theory, The Grainger College of Engineering,\and\small University of Illinois Urbana-Champaign, Urbana, Illinois 61801, USA\and
        \small$^{17}$Rice Laboratory for Emergent Magnetic Materials, Rice University, Houston, TX 77005, USA\and
	\small$^\ast$Corresponding authors. Emails: mingyi@rice.edu, bjalan@umn.edu, milan.radovic@psi.ch\and
	\small$^\dagger$These authors contributed equally to this work.
}

\begin{document} 

\maketitle

\begin{abstract} \bfseries \boldmath
Recently, rutile ruthenium dioxide (RuO$_2$) has attracted renewed interest due to expectations of prominent altermagnetic spin splitting. However, accumulating experimental evidence suggests that, in its bulk and thick-film forms, RuO$_2$ does not display any form of magnetic ordering. Despite this, the spin structure of RuO$_2$ remains largely unexplored in the ultrathin limit, where substrate-imposed epitaxial strain can be substantial. Here, we use spin-resolved angle-resolved photoemission spectroscopy, supported by ab initio calculations, to reveal the electronic structure of 2-nanometer-thick epitaxial RuO$_2$ heterostructures. We observe an unconventional spin texture characterized by the coexistence of mirror-even and mirror-odd momentum-dependent components. A comprehensive symmetry analysis rules out nonmagnetic origins of this spin texture. These findings suggest an emergent nonrelativistic spin structure enabled by epitaxial strain in the ultrathin limit, marking a distinct departure from the behavior of relaxed or bulk RuO$_2$. Our work opens previously unexplored perspectives for exploring symmetry-breaking mechanisms and spin textures in oxide heterostructures.
\end{abstract}

\noindent
\subsection*{Introduction}
Conventional ferromagnetism and antiferromagnetism are often considered the two archetypal classes of collinear magnetic orders. Although the former displays a net magnetization, in a collinear N\'eel antiferromagnet opposite spins are located in sublattices that are related by a lattice translation. However, sublattices with opposite spins can also be related by point-group symmetries of the lattice. Recently, a theoretical classification of collinear magnetic phases based on spin-group theory revealed another type of collinear magnetic order, altermagnetism, where the magnetic sublattices with opposite spins are related by rotations~\cite{PhysRevX.12.031042,PhysRevX.12.040501}. The hallmark of such a fully-compensated magnetic order in the electronic spectra is a nodal nonrelativistic spin splitting with $d$-, $g$-, or $i$-wave symmetry~\cite{Nodal2024}. This is, on symmetry grounds, the same spin splitting realized in metals undergoing a spin-triplet even-parity Pomeranchuk-type instability~\cite{PhysRevLett.93.036403,PhysRevB.75.115103}, although altermagnets can also be insulators. Beyond collinear spin arrangements, the spin group formalism has also been recently applied in the classification of noncollinear coplanar and noncoplanar magnetic orders~\cite{PhysRevX.14.031037,PhysRevX.14.031038,PhysRevX.14.031039}, revealing unusual odd-parity magnetic phases \cite{Hellenes2023}. Overall, the studies of altermagnets and odd-parity magnets revealed a plethora of unique phenomena including electronic topology~\cite{PhysRevX.12.021016,PhysRevLett.134.096703,Gao2023}, unconventional spin textures~\cite{Fernandes2024,PhysRevX.14.031037,Hu2025,Radaelli2025}, and chiral responses~\cite{PhysRevLett.131.256703,PhysRevLett.133.156702,CrSb_chiral_DFT,Gao2023}. Such underlying physics harbored by complex unconventional magnets provides a fertile playground of experimenting with various types of Hall responses~\cite{Libor2020,PhysRevX.14.031038,PhysRevLett.133.086503,Takahashi2025} and spin-charge conversion~\cite{Hu2025,Gonzalez-Hernandez2024}, inspiring the development of next-generation spintronics, multiferroics, and thermoelectrics~\cite{PhysRevX.12.040501,PhysRevX.12.011028,DalDin2024,Liu2025,PhysRevB.110.094427,Jungwirth2025,Libor_multiferroics}.

Among the predicted altermagnets, RuO$_2$ was considered one of the most promising candidates for applications for being a metal with room temperature magnetic order, owing to earlier experimental reports supporting magnetism in these compounds~\cite{PhysRevLett.122.017202,PhysRevLett.118.077201} and the large energy scale of altermagnetic spin splitting identified in ab-initio-based calculations~\cite{PhysRevB.99.184432,Libor2020,PhysRevX.12.040501,Guo2023,PhysRevX.12.011028}. In addition, field-induced nonlinear Hall effect~\cite{Feng2022,Tschirner2023}, planar Hall effect~\cite{Song2025}, spin-splitting magnetoresistance effect~\cite{Chen2025}, spin-splitter torque effect (charge-to-spin)~\cite{PhysRevLett.128.197202,PhysRevLett.129.137201,Bose2022,Zhang_torque_2024,PhysRevLett.132.086701}, and efficient spin-to-charge conversion~\cite{PhysRevLett.130.216701,PhysRevLett.133.056701} were experimentally reported in RuO$_2$ films, the origin of which was attributed to the underlying $d$-wave altermagnetism. Angle-resolved photoemission spectroscopy (ARPES) measurements employing magnetic circular dichroism on strain-relaxed RuO$_2$ films has reported the observation of time-reversal-symmetry (TRS) breaking~\cite{Fedchenko2024,Lytvynenko2026}. One spin-resolved ARPES work has reported a $d$-wave spin texture in RuO$_2$ single crystals~\cite{Lin2024}. On the other hand, several experimental probes including x-ray diffraction, unpolarized and polarized neutron diffraction, and muon spin rotation spectroscopy have ruled out the presence of altermagnetism (or any types of magnetism) in both bulk and strain-relaxed film forms of RuO$_2$~\cite{Kebler2024,PhysRevLett.132.166702,Kiefer2025}. The absence of altermagnetism in bulk RuO$_2$ single crystals was further supported by infrared spectroscopy~\cite{PhysRevB.111.L041115}, quantum oscillations~\cite{Wu2025}, torque magnetometry and magnetization measurements~\cite{Qian2025}, as well as Mössbaeur spectroscopy, nuclear forward scattering, and inelastic x-ray and neutron scattering~\cite{Yumnam2025}. Moreover, in the thin film limit, detailed transport examination in ferromagnetic/RuO$_2$ heterostructures reported the absence of altermagnetic spin splitting down to 5~nm of RuO$_2$ thickness~\cite{Wang2026}. This is consistent with the results from time-domain terahertz spectroscopy~\cite{Plouff2025}. Recent ARPES and spin-resolved ARPES measurements on both single crystals and strain-relaxed films of RuO$_2$ have also reached the nonmagnetic conclusion based on measurements of the in-plane spin polarization~\cite{PhysRevLett.133.176401,Osumi2026}. Theoretical reexamination based on first-principles concomitantly revealed the fragile nature and delicate phase boundary of magnetism in RuO$_2$~\cite{PhysRevB.109.134424,Qian2025_DFT}.

Despite intense debates and a convergence towards the nonmagnetic nature of RuO$_2$ in bulk and thin film forms, the ultrathin epitaxially strained regime of RuO$_2$ below a critical thickness of around 4~nm is rarely explored and far from completely understood. Experimentally, it was demonstrated that hybrid molecular beam epitaxy (hMBE)-grown RuO$_2$ below 4~nm in RuO$_2$/TiO$_2$(110) films is fully strained by the substrate along both in-plane directions, reaching -4.7\% compressive strain along [001] and +2.3\% tensile strain along [1$\bar{1}$0]\cite{Jeong2025}. However, in sputter-grown RuO$_2$ thin films, metallic behavior has not been reported below this thickness limit. Instead, hMBE-grown RuO$_2$ thin films on TiO$_2$ (110) substrates exhibit metallic behavior within the fully strained ultrathin thickness regime, attributed to excellent stoichiometry control and atomically smooth interfaces and surfaces~\cite{PhysRevMaterials.8.075002,Jeong2025}. Moreover, structural characterization yields a polar structure with point group $mm2$ ($C_{2v}$)~\cite{Jeong2025}. Importantly, rotational-anisotropy second-harmonic generation and magneto-optical measurements suggested a TRS-broken altermagnetic polar metallic phase in epitaxially strained RuO$_2$ with moments perpendicular to the film direction ~\cite{Jeong2025,Weber2024}, accompanied by the observation of unusual nonlinear Hall signals at low fields only for the thin and fully strained RuO$_2$ layers~\cite{Jeong_AHE_PNAS}. These are in contrast with the observations in thin-to-thick films and single crystals of RuO$_2$. In particular, both experimental results and theoretical calculations suggest that epitaxial strain plays a crucial role in stabilizing the magnetic states of RuO$_2$ \cite{Jeong2025,Jeong_AHE_PNAS,Meinert2025}. Therefore, spectroscopy investigation with spin and momentum resolution to resolve the electronic band dispersion of ultrathin epitaxially strained RuO$_2$ and its spin character is highly desired.

Here, using spin-resolved ARPES, we probe the momentum-spin-energy resolved one-electron spectral density of the 2~nm epitaxially strained metallic RuO$_2$ films grown by hMBE, a regime of dimensionality and strain that has not been achieved before in spin-resolved ARPES experiments. Narrow bands arising from surface states in proximity to the Fermi level modified by the epitaxial strain are observed. Furthermore, we unveil the coexistence of mirror-odd and mirror-even $k$-space spin texture by measuring both in-plane and out-of-plane photoelectron spin polarization. Through symmetry analyses and numerical simulations of the spin-polarized spectra, we find that such observed spin polarization cannot be explained by the polar character of the crystal nor by photoemission extrinsic effects, and must point to the intrinsically broken TRS in ultrathin epitaxially strained RuO$_2$ films. The possible magnetic point groups are deduced theoretically, consistent with both a ferromagnetic phase and a $d$-wave altermagnetic phase with magnetic moments pointing within the film plane.

\subsection*{Results}
\begin{figure}
    \centering
    \includegraphics[width=0.65\linewidth]{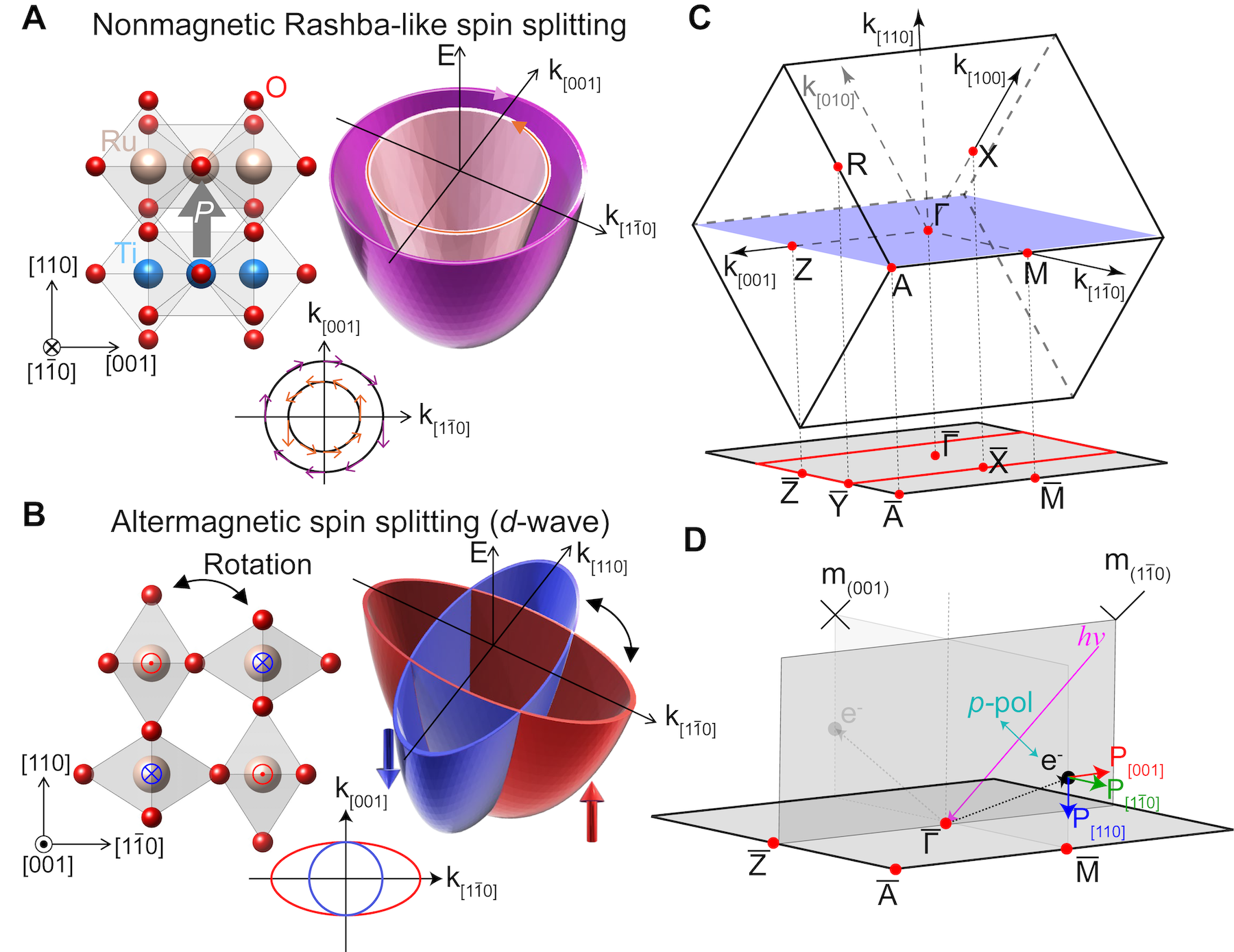}
    \caption{\textbf{Proposed spin texture relevant in epitaxially strained RuO$_2$ and a spin-resolved photoemission measurement geometry.} (\textbf{A}) Schematic illustration of the charge dipole produced at the (110) RuO$_2$/TiO$_2$ interface and the associated Rashba-type spin splitting within the (110)-plane. (\textbf{B}) Decorated local chemical environment of the two Ru sublattices in RuO$_2$ related by a C$_4$ rotational symmetry and the schematic illustration of nonrelativistic altermagnetic spin splitting in $k$-space. (\textbf{C}) Bulk Brillouin zone (BZ), its projection along [110] (light blue and gray planes) and the (110) surface Brillouin zone (red rectangle) of RuO$_2$. (\textbf{D}) A schematic illustration of one photoemission geometry employed in our experiments, termed as ``Geometry A'' in Table.~\ref{tab:sarpes}. The (1$\bar{1}$0)-mirror of the total photoemission system is indicated as preserved, while the (001)-mirror is indicated as broken. The three photoelectron spin components are indicated in red, green, and blue arrows.}
    \label{fig:fig1}
\end{figure}

Given the RuO$_2$/TiO$_2$ (110) heterostructure, we begin with an illustration of the possible mechanisms that could give rise to spin textures in this geometry. A schematic illustration of the RuO$_2$/TiO$_2$ interface is given in Fig.~\ref{fig:fig1}A, where an electric polarization $\textbf{P}$ can be formed along the [110] axis, breaking the inversion symmetry of the system. Consequently, relativistic spin splitting can appear, with an example of the Rashba-type spin texture shown in the plane spanned by the [001] and [1$\bar{1}$0] axes (i.e., the film plane). Such spin texture would show mirror-odd behavior about any vertical mirrors. Meanwhile, if altermagnetism exists between the two Ru sublattices characterized by $[C_2||C_{4z}\textbf{{t}}]$ symmetries of the bulk $4/mmm$ point group (here the $z$ axis refers to the [001] in bulk geometry and $\mathbf{t}$ to a half-translation), as shown in Fig.~\ref{fig:fig1}B, a $d$-wave spin splitting pattern about the [001] axis would emerge, giving rise to spin-splitting that is even with respect to the (001) and (1$\bar{1}$0) mirrors. Going from real space to momentum space, we show in Fig.~\ref{fig:fig1}C the Brillouin zone of a strained-RuO$_2$-structure with the [110] $k$-axis pointing along $z$ as the experimental out-of-plane direction. Hence, the ARPES in-plane measurements are carried out within the $\Gamma$-$M$-$A$-$Z$ plane projected to the surface. 

An important point that needs to be taken into account in interpreting spin-resolved ARPES measurements is that the combination of incident linearly-polarized light and spin-orbit coupling (SOC) can give rise to spin-splittings of bands through Rashba and other related effects~\cite{yang2026SOC}. Moreover, extrinsic effects related to the geometry of the experiment and to the detection of photoelectrons may result in an apparent spin-polarized spectrum even when the crystal does not intrinsically break TRS. Multiple scattering of spin-orbit coupled photoelectrons can produce phase shift and additional spin polarizations in the final states beyond initial states under specific geometries, not only for circularly polarized photons, but also under linearly polarized or even unpolarized light~\cite{PhysRevLett.59.934,Tamura_1991,Henk_1994,PhysRevLett.60.651,PhysRevB.45.3849,IRMER1996321,FEDER1981547,Heinzmann_2012}. To illustrate how these extrinsic, measurement-related sources of spin polarization can affect our measurements, consider our measurement geometry in Fig.~\ref{fig:fig1}D (termed as Geometry A), where $p$-polarized photons shine on a RuO$_2$/TiO$_2$ (110) heterostructure (structure point group $mm2$) within the (1$\bar{1}$0) mirror plane at off-normal incidence. For an $mm2$ crystal without intrinsic TRS breaking, the detected photoelectron can still exhibit a net spin polarization perpendicular to the preserved mirror $m_{(1\bar{1}0)}$, which we denote $P_{[1\bar{1}0]}$ (green) in this case. Here we use $P$ to denote detected photoelectron spin polarization to distinguish from the intrinsic spin polarization of the initial Bloch states, $S$. This means in this geometry, even the observation of a finite $P_{[1\bar{1}0]}$ at normal emission ($\Gamma$) is allowed without intrinsic TRS breaking of the initial state, as demonstrated in previous studies~\cite{Henk_1994,IRMER1996321} and confirmed by our ab-initio-based one-step model spin-resolved ARPES calculations shown in the Supplementary Text (fig.~\ref{fig:fig_one_step}). In contrast, for the two spin components parallel to the (1$\bar{1}$0) mirror ($P_{[001]}$ and $P_{[110]}$ colored in red and blue, respectively), the net photoelectron spin polarization cannot be non-zero and must, therefore reverse its sign after the (1$\bar{1}$0)-mirror reflection. Furthermore, due to the beam path and photon polarization depicted in Fig.~\ref{fig:fig1}D, the crystal mirror $m_{(001)}$ is explicitly broken by the measurement geometry. Hence, the even or odd behavior of the photoelectrons with respect to the (001)-mirror cannot be inferred in this geometry. Instead, we can employ a different geometry (Geometry B) to probe for intrinsic spin texture associated with the $m_{(001)}$ mirror. A schematic illustration for Geometry B is shown in fig.~\ref{fig:fig_geometry_B} where the light incidence is rotated azimuthally 90$^\circ$ from Geometry A. The final state analysis here is reminiscent of the photoemission matrix elements where light with different linear polarizations can be used to probe orbitals with different symmetries even for a system with preserved orbital degeneracy.

To probe for intrinsic broken TRS in the initial state, we first list the expected final state selection rules in Table.~\ref{tab:sarpes} based on such analysis for an assumed TRS-invariant $mm2$ state of the epitaxially strained RuO$_2$ films measured in Geometry A and B. We have confirmed such final state selection rules numerically using the ab-initio-based fully-relativistic one-step model approach as shown in fig.~\ref{fig:fig_one_step}. Then any observed photoelectron spin polarization that violates these rules would indicate symmetry breakings beyond extrinsic effects, which in turn can be associated with intrinsic TRS breaking under our controlled experiments. In our following spin-resolved ARPES measurements, we carefully choose our measurement geometries, and key observations are made for $P_{[001]}$ in Geometry A and $P_{[110]}$ in Geometry B to examine the possibilities of magnetism in the epitaxially strained RuO$_2$.

To experimentally probe the spin texture, we synthesized fully strained metallic RuO$_2$ heterostructure via hMBE for spin-resolved ARPES measurements. To avoid charging effects during ARPES measurements, we designed an epitaxial RuO$_2$ heterostructure consisting of a nominal 2~nm RuO$_2$ layer and a 2~nm TiO$_2$ buffer layer grown on a conductive Nb:TiO$_2$ (110) substrate, as schematically shown in the inset of Fig.~\ref{fig:fig2}A. The conductive substrate effectively eliminated charging effects, while the TiO$_2$ buffer layer was adopted to suppress unintended interfacial effects due to surface contamination between the Nb:TiO$_2$ and the RuO$_2$ layer. Figure~\ref{fig:fig2} (A and B) show the high-resolution X-ray reflectivity (XRR) and diffraction (XRD) $\theta$-2$\theta$ scan results, confirming the (110)-oriented out-of-plane lattice structure of the RuO$_2$ heterostructures. The XRR data exhibit Kiessig fringes extending to $\sim$8$^\circ$, indicating atomically smooth surfaces, and fitting results (solid lines) confirm a RuO$_2$ thickness of 2.7~nm (Fig.~\ref{fig:fig2}A). XRD scans show clear Laue oscillations around the RuO$_2$ (110) Bragg peaks, indicating sharp interfaces. Reflection high-energy electron diffraction (RHEED) images in Fig.~\ref{fig:fig2}C taken after growth show streaky RHEED patterns accompanied by the Kikuchi lines along the [1$\bar{1}$0] crystal directions, indicating the excellent crystallinity. Atomic force microscopy (AFM) images in Fig.~\ref{fig:fig2}D further reveal an atomically smooth surface with a root mean square roughness ($S_q$) of 171.8~pm. Additionally, rotational anisotropy second-harmonic generation (RA-SHG) measurements (Fig.~\ref{fig:fig2}E) confirm the \textit{mm}2 point group symmetry, indicating a fully strained, noncentrosymmetric RuO$_2$ heterostructure~\cite{Jeong2025} (see fig.~\ref{fig:fig_SHG_more} for more details about the SHG analysis). Moreover, to confirm the surface stoichiometry of the RuO$_2$ heterostructures grown on TiO$_2$ and Nb:TiO$_2$ substrates, we have performed ex situ x-ray photoelectron spectroscopy (XPS) (fig.~\ref{fig:fig_XPS}), ambient pressure-XPS  (AP-XPS) (fig.~\ref{fig:fig_APXPS}), and in situ XPS measurements within the spin-resolved ARPES chamber (fig.~\ref{fig:fig_insitu_XPS}) to reveal the consistent Ru(IV) oxidation states across samples and the suppression of carbon-related contamination through our in situ oxygen annealing procedures (see Materials and Methods), in agreement with our previous RuO$_2$ films~\cite{Jeong2025,William_2021}.

Before we introduce the observed spin texture, we first present the key features in the electronic structure of the ultrathin epitaxially strained RuO$_2$ measured by spin-integrated ARPES. Similar to previously reported band structure of bulk and thick films of RuO$_2$~\cite{Jovic2018,Ruf2021,PhysRevLett.133.176401,Jovic2021}, the Fermi surface (FS) consists of a large $\Bar{\Gamma}$-centered hexagonal-like pocket surrounded by smaller pockets, although the fine details differ. However, the most distinguishable spectral features in the ultrathin epitaxially strained RuO$_2$ are the narrow bands (NBs) near the Fermi level that are largely dispersionless along the [1$\Bar{1}$0] axis. Specifically, two sets of NBs appear, one along $\Bar{\Gamma}-\Bar{M}$ ($\alpha$-NBs) and one near the zone boundary $\Bar{Z}-\Bar{A}$ ($\beta$-NBs), as can be observed on both constant energy contours (Fig.~\ref{fig:fig3} (A and B)) and high symmetry momentum slices (Fig.~\ref{fig:fig3}(C and E)). Both sets of NBs contribute to prominent peaks in the energy distribution curves (EDCs) integrated across the whole momentum ranges (Fig.~\ref{fig:fig3}(C and E)). The $\alpha$-NBs also exhibit resolvable wiggling along $\Bar{\Gamma}-\Bar{M}$, as evident in the distinct peak positions for the single EDC extracted at $\Bar{\Gamma}$ compared to the integrated EDC (Fig.~\ref{fig:fig3}C). We note that the $\alpha$-NBs appear at 120~meV below \ef, distinct from that of the surface narrow bands observed within 50~meV below \ef~in cleaved single crystals~\cite{Jovic2018,PhysRevLett.133.176401}. They are also not observed in our thicker 14~nm films grown by the same methods but under surface annealing at higher temperatures (see Materials and Methods for details) prior to ARPES measurements (fig.~\ref{fig:fig_14nm_ARPES}), nor present in other strain-relaxed films in previous literature~\cite{Fedchenko2024,PhysRevLett.133.176401,Ruf2021}. Such discrepancy may be attributed to different surface stoichiometries obtained through different post-annealing temperatures, as the $\alpha$-NBs-analogous feature has been shown to disappear on a Ru-rich surface (with oxygen vacancies)~\cite{Jovic2021}.

To gain insights into the origins of the NBs, we first carried out spin-polarized bulk density functional theory (DFT) calculations. Having explored the effects of the full epitaxial strain and Hubbard $U$ correction, we arrived at the best match between calculated and measured dispersions with no $U$ but unrelaxed strain where the system hosts a ground state of uncompensated altermagnetism, as shown in Fig.~\ref{fig:fig3} (C and E) (see Supplementary Text and fig.~\ref{fig:fig_bulk_DFT} for full details). Notably, the $\beta$-NBs observed at about 300~meV below \ef~in Fig.~\ref{fig:fig3}E have been discussed in previous literature proposing strain-stabilized superconductivity in RuO$_2$ with ARPES measurements down to 7~nm where partial strain relaxation exists~\cite{Ruf2021}. Our bulk DFT calculations confirm such a strain-tuning trend towards \ef~for the $\beta$-NBs and our ARPES data observe it in the fully strained regime of the 2.7~nm RuO$_2$. However, the $\alpha$-NBs are still not captured by these calculations, as shown in Fig.~\ref{fig:fig3}C. Therefore, we have carried out DFT slab calculations of RuO$_2$ fully strained by TiO$_2$ under no Hubbard U correction (see fig.~\ref{fig:fig_nosp_slab_1}, fig.~\ref{fig:fig_nosp_slab_2}, and Supplementary Text). Due to the delicate nature of the magnetic ground states in RuO$_2$ slab calculations, we do not rely on them as an indicator for magnetic orders. Therefore, non-spin-polarized calculations are shown in Fig.~\ref{fig:fig3} (D and F) only in assistance of understanding the orbital components of the NBs. As evidenced by the direct comparison between Fig.~\ref{fig:fig3} (C and D), all three major band features ($\alpha$-NBs, $\gamma$, and $\delta$) observed by ARPES are nicely reproduced by the slab calculations after a shift of the DFT \ef~by -200 meV. In particular, even the wiggling of the $\alpha$-NBs is captured in Fig.~\ref{fig:fig3}D. It is important to emphasize that such an \ef~adjustment is not a deliberate choice to only fit the $\alpha$-NBs, but a rigid shift to match all the bands, which can be further validated by the $\beta$-NBs comparison in Fig.~\ref{fig:fig3} (E and F) involving consistent ARPES data, bulk DFT, and slab DFT calculations, after the same \ef~shift applied to the slab calculations. To comply with the surface sensitivity of the vacuum ultraviolet ARPES, the marker size and transparency of the slab calculations correspond to the projected weights onto the summed Ru $d$ and O $p$ orbitals within the top two surface layers of RuO$_2$. Through further partial charge projection on Ru $d$ orbitals, it is clear that the $\alpha$-NBs have strong Ru $d_{z^2}$ (red) characters near the surface. This feature, combined with the absence of the $\alpha$-NBs on the bulk calculation, suggests that its origin are the less-hybridized surface Ru $d_{z^2}$ orbitals (see fig.~\ref{fig:fig_nosp_slab_2}) that ``stick out'' of the surface. The fact that this orbital points out of plane would naturally explain the weak in-plane dispersion of this band, in agreement with our ARPES observations. Notice that half of the thickness of the constructed slab exceeds 2~nm, which is close to the experimental thickness of our RuO$_2$ films. Hence, combining both the experimental observation and theoretical calculations, we conclude that the ARPES observed $\alpha$-NBs, reminiscent of the narrow band surface states reported in single crystals, are distinct in several aspects. First, their presence at around $E-$~\ef~$\approx-120$~meV is an epitaxial-strain-driven behavior, absent in strain-relaxed films and single crystals. Second, by pushing the film thickness close to the surface limit, the $\alpha$-NBs largely remain across the film (see fig.~\ref{fig:fig_nosp_slab_2} and Supplementary Text for full details). Third, this band likely arises from surface rather than bulk states. The impact of these narrow bands on the electronic properties of ultrathin RuO$_2$ films, including its possible role in aiding electronic instabilities~\cite{Kebler2025}, deserves further theoretical investigations that are beyond the scope of our present work. Overall, the spin-integrated ARPES data show excellent agreement with slab DFT calculations assuming stoichiometric terminations, and the valence-band behavior at deeper binding energies further supports that the epitaxially strained RuO$_2$ films maintain a near-stoichiometric surface~\cite{Jovic2021} (see fig.~\ref{fig:fig_deeperE_ARPES} and Supplementary Text). This is fully consistent with the conclusions drawn from the in situ XPS measurements (fig.~\ref{fig:fig_insitu_XPS}). 

Next, we examine the in-plane spin texture $P_{[001]}$ with respect to the (1$\bar{1}$0)-mirror of the epitaxially strained RuO$_2$ film as measured by spin-resolved ARPES. We emphasize that Fig.~\ref{fig:fig4} derives from measurements in Geometry A (see Fig.~\ref{fig:fig1}D and Table.~\ref{tab:sarpes}) and the sample was magnetized in situ by an in-plane magnetic field of 0.2~T pointing 45$^\circ$ between the [001] and the [1$\bar{1}$0] film axes. A Fermi surface (FS) measured by 62~eV photons is shown in Fig.~\ref{fig:fig4}A for navigating the spin-resolved EDCs. The measured band dispersions along $\bar{\Gamma}$-$\bar{M}$ (dashed orange double arrow), which is perpendicular to the mirror, are shown in Fig.~\ref{fig:fig4}B. They show similar features to the 62 eV $\bar{\Gamma}$-$\bar{M}$ data in Fig.~\ref{fig:fig3}C measured under Geometry B, but differ in terms of intensity asymmetry between $+k_{[1\bar{1}0]}$ and $-k_{[1\bar{1}0]}$ due to the photoemission matrix element effects. To probe the spin polarization of bands along $\bar{\Gamma}$-$\bar{M}$, we measure spin-resolved EDCs at momenta denoted by the magenta and cross symbols in Fig.~\ref{fig:fig4} (A and B). At finite momenta ($k_{\rm [1\bar{1}0]}\approx\pm 0.5\textup{~\AA}^{-1}$), the measured spin-resolved EDCs are displayed in Fig.~\ref{fig:fig4} (C and D), where spin up and down are respectively represented by red and blue curves and fine details are presented in the zoom-in panels emphasizing the differences between the spin up and down photoelectron counts. Consequently, the derived spin polarization in Fig.~\ref{fig:fig4} (E and F) shows a (1$\bar{1}$0)-mirror-odd (flipped spins upon reversing $k_{\rm [1\bar{1}0]}$) behavior for binding energies away from the Fermi level, consistent with the final state selection rules in Table.~\ref{tab:sarpes}. However, near the Fermi level, a (1$\bar{1}$0)-mirror-even (spins not flipped upon reversing $k_{\rm [1\bar{1}0]}$) behavior arises, which cannot be explained by the final state effects of an assumed TRS-preserved state. Furthermore, we directly measure the [001] spin polarization at normal emission ($k_{//}=0$). If the RuO$_2$ film had preserved TRS, the $P_{[001]}$ at $\Gamma$ would be enforced to be zero according to the selection rules of Table.~\ref{tab:sarpes}. Remarkably however, as shown in Fig.~\ref{fig:fig4} (G and H), a net spin polarization of $P_{[001]}$ is observed. This also means that the $\alpha$-NBs are spin-polarized. Photon energy dependent counterparts of Fig.~\ref{fig:fig4} (C to H) are presented in fig.~\ref{fig:fig_geometry_A_GM_S001_hv_dep} and fig.~\ref{fig:fig_geometry_A_G_S001_hv_dep} showing a complicated evolution due to multiple scattering of photoelectrons but a consistent symmetry-breaking (1$\bar{1}$0)-mirror-even spin polarization component. These observations suggest that the measured epitaxially strained RuO$_2$ film exhibits intrinsic magnetic spin polarization, indicating broken TRS. Regarding the out-of-plane spin polarization on the $\bar{\Gamma}$-$\bar{M}$ path,  $P_{[110]}$, one set of measurements under Geometry B presented in fig.~\ref{fig:fig_Geometry_B_GM_55eV_S110} suggests a near-zero amplitude compared to its in-plane components. In addition, measurements on the $\Bar{\Gamma}$-$\Bar{Z}$ path and along the diagonal in-plane direction for $P_{[001]}$ are performed in Geometry A, as shown in fig.~\ref{fig:fig_geometry_A_GZ_diag_62eV_S001}. However, due to the broken (001)-mirror of the total photoemission system, we have observed a complicated mixture of (001)-mirror-even and odd behavior. The same argument applies as well to the coexisting (1$\bar{1}$0)-mirror-odd and even $P_{[001]}$ spin polarization on the $\bar{\Gamma}$-$\bar{M}$ path measured alternatively under Geometry B in fig.~\ref{fig:fig_old_main_4}, where the (1$\bar{1}$0)-mirror is no longer a preserved mirror of the total photoemission system.

Finally, we investigate the [110] out-of-plane spin polarization to show that $P_{[110]}$ channel obeys the $mm2$ TRS-preserving final state selection rules (Table.~\ref{tab:sarpes}), indicating that intrinsic symmetry breakings are only unambiguously observed for in-plane spins. In Fig.~\ref{fig:fig5}, the sample was magnetized in situ by an even stronger out-of-plane magnetic field of 0.4~T and Geometry B was chosen. Nonetheless as indicated by the selection rules in Table.~\ref{tab:sarpes}, $P_{[110]}$ would flip sign with respect to the preserved mirror plane of the total photoemission system regardless of Geometry A or B under the TRS-preserved assumption. In Fig.~\ref{fig:fig5} (A and B), we show the measured FS and $\Bar{\Gamma}$-$\Bar{Z}$ band dispersions using 55~eV $p$-polarized light. The differences compared to ARPES spectra measured at 62~eV in Fig.~\ref{fig:fig3}A and fig.~\ref{fig:fig_Geometry_B_GZ_G_62eV_55eV_S110_S001} (A and B) must be due to kinetic-energy-dependent photoemission matrix element effect rather than $k_z$ dispersions, as the ultrathin film does not have a well-defined $k_z$. Spin-resolved EDCs probing $P_{[110]}$ at finite momenta with respect to the preserved (001)-mirror are presented in Fig.~\ref{fig:fig5} (C and D), and the converted spin polarization in Fig.~\ref{fig:fig5} (E and F). We observe a pure (001)-mirror-odd behavior that is compatible within the TRS-preserved final state selection rules (Table.~\ref{tab:sarpes}). The comprehensive check of the mirror-odd behavior of the out-of-plane spin $P_{[110]}$ encompassing measurements using different photon energies and at both finite and zero momenta are presented in detail in fig.~\ref{fig:fig_Geometry_B_GZ_G_62eV_55eV_S110_S001} and the Supplementary Text. Besides the out-of-plane $P_{[110]}$ and the in-plane $P_{[001]}$, the third spin quantization axis [1$\bar{1}$0] has also been selectively measured under Geometry B and after the same experimental preparation as in Fig.~\ref{fig:fig4}, as detailed in Supplementary Text and fig.~\ref{fig:fig_geometry_B_GZ_G_S1-10}, showing dominantly (001)-mirror-odd behavior but a small even component close to experimental error bars. This violates the final state selection rules tabulated in Table.~\ref{tab:sarpes} which enforce $P_{[1\bar{1}0]}$ to flip sign after the (001)-mirror reflection. Therefore, we conclude that the TRS-broken states at 15~K in the 2~nm ultrathin epitaxially strained RuO$_2$ films originate from electrons with predominantly in-plane, rather than out-of-plane, spin texture.

\subsection*{Discussion}
Our observations establish a momentum-dependent coexisting mirror-odd and mirror-even $k$-space spin texture, which we associate with broken TRS in the ultrathin epitaxially strained RuO$_2$/TiO$_2$ (110) heterostructure, after comprehensively ruling out nonmagnetic origins. This is consistent with suggestions from other experimental probes, including second-harmonic generation, magneto-optic Kerr effect (MOKE), time-resolved MOKE, and Hall transport~\cite{Jeong2025,Weber2024,Jeong_AHE_PNAS}. In addition, one recent work~\cite{Akashdeep2026} employing depth-resolved low-energy muon spin rotation/relaxation measurements suggests a possible reconciliation with nonmagnetic claims of RuO$_2$ films of tens of nm thickness, where magnetic signals are claimed to be observed near the surface region. Here, by pushing RuO$_2$ films towards 2~nm and the fully epitaxially strained regime, it appears that magnetism can be promoted, consistent with recent polarized neutron reflectometry conclusions~\cite{Jeong2026_PNR}.

It is interesting to note that the paramagnetic point group $mm2$ of the strained ultrathin RuO$_2$ films is polar, and as such can host unusual topological phenomena. For instance, Kramers nodal lines could exist along the [110] direction, protected by the (001) and (1$\bar{1}$0) mirrors \cite{Xie2021}. Such Kramers nodal lines can support distinct spin textures, as previously reported in SmAlSi and intercalated transition metal dichalcogenides~\cite{Zhang2023_2,Zhang2025,domaine2025}. However, in our experiment, $k_{110}$ is not a good quantum number, since this is the out-of-plane direction of the ultrathin film.

While the analysis above compared the experimental data with the final-state spin-resolved ARPES selection rules, we now perform a group-theory analysis to narrow down the magnetic point groups consistent with the observed spin texture. Assuming the system at high temperatures belongs to the paramagnetic point group $mm2.1' $, we classify the spin-splitting terms that are linear in spin and up to quadratic order in momentum in terms of the irreducible representations (irreps) of the point group, see Table.~\ref{tab:irreps}; other details are explained in the Supplementary Text and table.~\ref{tab:mm2_table}. While symmetry alone cannot predict the magnitude of the spin splittings, it serves as a tool to infer the symmetry properties of the magnetic order parameter. Specifically, we search for the simplest combinations of irreps that give the observed spin-splitting terms in both geometries. In this sense, our analysis gives the highest-symmetry magnetic point group consistent with the data, but lower-symmetry groups cannot be ruled out. In Geometry A, three spin-splitting terms are observed: the $k$-odd term $k_{1\bar{1}0}\sigma_{001}$ and the $k$-even term $k_{1\bar{1}0}^2\sigma_{001}$ in Fig. \ref{fig:fig4} (E and F) as well as the constant $\sigma_{001}$ term in Fig. \ref{fig:fig4}H. The $k$-odd term, according to Table.~\ref{tab:irreps}, can be ascribed to the $A_1^+$ irrep, which is to say that it is due to the intrinsic symmetry of the film without the magnetic order. In this notation, the superscript denotes a time-reversal even ($+$) or odd ($-$) irrep. In particular, it is a Rashba-like SOC term that is present due to the inversion-breaking effect of the film surface. The $k$-even terms, however, break the time reversal symmetry and transform as the $B_1^-$ irrep. We propose that this is the primary magnetic order parameter in our film, consistent with the discussion based on the final-state selection rules. In Table \ref{tab:irreps}, we highlight with underlines the three terms identified in Geometry A. Still in what concerns Geometry A, fig. \ref{fig:fig_geometry_A_GZ_diag_62eV_S001} (H to I) shows a $k$-even term $k_{001}^2\sigma_{001}$ that is also consistent with the $B_1^-$ irrep, although an odd-in-$k$ term inconsistent with this irrep is also seen. We note, however, that the experiment in Geometry A breaks the mirror plane perpendicular to $k_{001}$, which makes the interpretation of these terms less clear. For this reason, we do not highlight these terms in Table \ref{tab:irreps}.

Interestingly, there are spin-splitting terms observed in Geometry B that belong to different irreps than those observed in Geometry A, attesting to the fact that measurements in different geometries can give complementary information. Starting with the spin polarization along the $[110]$ direction, whose corresponding data are shown in Fig.~\ref{fig:fig5} and fig.~\ref{fig:fig_Geometry_B_GZ_G_62eV_55eV_S110_S001} (E, F, and L), we observe an odd-in-$k$ term (but no even-in-$k$ or constant term) of the form $k_{001}\sigma_{110}$. This term transforms as the $B_1^+$ irrep, which breaks the same mirror symmetry as that broken by Geometry B (table.~\ref{tab:mm2_table}). Hence, the $k_{001}\sigma_{110}$ term can be interpreted as a SOC-effect that does not require TRS breaking. This term is highlighted in bold in Table \ref{tab:irreps}. For measurements taken along the momentum $k_{1\bar{1}0}$ perpendicular to the mirror plane broken explicitly by Geometry B, fig. \ref{fig:fig_Geometry_B_GM_55eV_S110}, a negligible spin splitting is observed for the spin component along the $[110]$ direction.

Moving on to the spin polarization along the $[001]$ direction in Geometry B, we observe a constant term $\sigma_{001}$ and an even-in-$k$ term $k_{001}^2\sigma_{001}$ in fig.~\ref{fig:fig_Geometry_B_GZ_G_62eV_55eV_S110_S001} (I, J, and N), which transforms as the $B_1^-$ irrep, as highlighted in bold in Table \ref{tab:irreps}. While this is a TRS-odd irrep that agrees with the TRS-odd irrep inferred from the measurements in Geometry A, such a term could also be explained by extrinsic final-state ARPES selection rules for Geometry B. Interestingly, fig. \ref{fig:fig_old_main_4} (I, J, K, L) suggests a combination of even-in-$k$ and odd-in-$k$ terms $k_{1\bar{1}0}^2\sigma_{001}$ and $k_{1\bar{1}0}\sigma_{001}$, which would transform as the irreps $B_1^-$ and $A_1^{+}$, respectively. However, because Geometry B explicitly breaks the mirror perpendicular to $k_{1\bar{1}0}$, these terms are not highlighted in Table \ref{tab:irreps}.

Finally, we analyze the data for the $(1\bar{1}0)$ spin polarization taken in Geometry B. Figure~\ref{fig:fig_geometry_B_GZ_G_S1-10} shows a dominant odd-in-$k$ term $k_{001}\sigma_{1\bar{1}0}$, which transforms as the trivial irrep $A_1^+$ and therefore is allowed by the lack of inversion symmetry of the crystal already in the nonmagnetic phase. The same figure also shows smaller contributions from a constant $\sigma_{1\bar{1}0}$ term and an even-in-$k$ term $k^2_{001}\sigma_{1\bar{1}0}$, both of which transform as the TR-odd irrep $B_2^-$. This is different from the TR-odd irrep $B_1^{-}$ obtained from the analysis of Geometry A, which cannot be explained by the final-state selection rules nor by the combined effect of the primary magnetic order and the explicit breaking of the $m_{(1\bar{1}0)}$ mirror by Geometry B, since $B_1^-\otimes B_1^+\neq B_2^-$. It is unclear, from our present investigations, whether this small term reflects another intrinsic magnetic order parameter or is an artifact due to tilting of magnetic moments, existence of minor competing micromagnetic domains, or depolarization fields that might be relevant in ultrathin film geometries. These three terms are also highlighted in bold in Table \ref{tab:irreps}. 

Thus, based on the combined final-state selection-rules analysis and intrinsic symmetry analysis, we propose that the observed spin-texture pattern is consistent with the condensation of a primary magnetic order parameter $B_1^-$, and thus with the magnetic point group $m'm2'$, with an additional weaker contribution from $B_2^-$, further reducing the symmetry to merely $2'$. This is consistent both with ferromagnetism and $d$-wave altermagnetism with in-plane moments. The latter scenario is consistent with the general first-principles expectation that RuO$_2$ is close to an altermagnetic instability, since in the presence of SOC, this type of altermagnetic order triggers weak ferromagnetism. We note that this magnetic group is different from that derived from previous SHG experiments in ultrathin films conducted near room temperature instead of 15~K, which found a $m'm'2$  magnetic point group~\cite{Jeong2025}. Although both correspond to $d$-wave altermagnetism, and in fact could arise from the same spin group in the absence of SOC, in $m'm2'$ the moments are in the plane of the film whereas for $m'm'2$, they point out-of-plane.

In summary, we have made the following key observations on 2~nm ultrathin epitaxially strained RuO$_2$ films: i) $\alpha$-NBs and $\beta$-NBs under the effect of strain, ii) mirror-even photoelectron spin polarization pointed along the [001] direction, and iii) mirror-odd photoelectron spin polarization. While iii) is a direct consequence of the inversion-symmetry breaking in the RuO$_2$/TiO$_2$ system, ii) is unique to the ultrathin epitaxially strained RuO$_2$ film not observed in bulk single crystals or thicker films. Their occurrence together here suggests a plausible mechanism where strain plays an essential role in stabilizing the altermagnetic phase, as discussed also in~\cite{Jeong2025}. It will be interesting to elucidate whether the NBs could also play a role in promoting this instability. Our study therefore reveals important roles played by strain and interfacial effects in RuO$_2$, and is instrumental for understanding the debated altermagnetic nature of RuO$_2$ and its further spintronic, optoelectronic, and electrocatalysis applications.



\subsection*{Materials and Methods}

\subsubsection*{Hybrid molecular beam epitaxy}

Epitaxial RuO\textsubscript{2} heterostructures composed of a 2.7~nm RuO\textsubscript{2} layer and a 2~nm TiO\textsubscript{2} buffer layer were grown on Nb:TiO\textsubscript{2} (110) single-crystal substrates (0.5~wt\% Nb, Crystec) using an oxide hybrid molecular beam epitaxy (hMBE) system (Scienta Omicron). Before growth, the substrates were cleaned with acetone, methanol, and isopropanol, followed by baking at 200~°C for 2~h in the load-lock chamber. Surface treatment was performed with an oxygen plasma anneal at 300~°C for 20~min to remove residual contaminants. RuO\textsubscript{2} layers were grown using a thermally evaporated metal-organic precursor, Ru(acac)\textsubscript{3}, from a low-temperature effusion cell (MBE Komponenten) maintained at 170-180~°C. The TiO\textsubscript{2} buffer layer was deposited using titanium tetraisopropoxide (TTIP, 99.999\%, Sigma-Aldrich) introduced through a gas inlet system at a beam equivalent pressure of $3\times10^{-7}$ Torr. Both layers were grown at a substrate temperature of 300~°C in an oxygen plasma environment (250~W RF power, $5\times10^{-6}$ Torr chamber pressure). After growth, the samples were cooled to 120~°C in the presence of oxygen plasma to suppress the formation of oxygen vacancies.

\subsubsection*{Optical second-harmonic generation}

To characterize the structural symmetry of 2.7~nm RuO$_2$/2~nm TiO$_2$/Nb:TiO$_2$ (110), we conducted rotational anisotropy second-harmonic generation (RA-SHG) measurements. An 800~nm femtosecond pulsed laser with a repetition rate of 80~MHz (VITARA-T, Coherent) was focused onto the sample with a beam size of about 20~$\mu$m at an oblique incidence of 45$^\circ$. The incident fundamental light was set to be either P- or S-polarized (P$_{\rm in}$ or S$_{\rm in}$), and the second-harmonic light from the sample was obtained in both P- or S-polarization (P$_{\rm out}$ or S$_{\rm out}$). To isolate the second-harmonic signal, the fundamental light was blocked using a combination of a 450~nm short-pass filter and a 400~nm band-pass filter (Thorlabs). The fundamental light was modulated by a mechanical chopper with a prime number frequency to suppress artifact signals. The modulated SHG signals obtained by the photomultiplier tube (Hammatsu) were demodulated by a lock-in amplifier (SR830, Stanford Research Systems).

\subsubsection*{Photoemission spectroscopy}

The ARPES and spin-resolved ARPES experiments on the 2~nm RuO$_2$/2~nm TiO$_2$/Nb:TiO$_2$ (110) substrate were performed at the Advanced Light Source, beamline 10.0.1.2. After transferring into the ultrahigh vacuum preparation chamber, the film was annealed at 295 °C under an oxygen-rich environment of $5\times10^{-6}$ Torr for 30 minutes. The oxygen pressure was maintained during the cooling down process towards room temperature. Upon transferring the films into the ARPES chamber and cooling down to a base temperature of around 15~K, a permanent magnet with an in-plane field of 0.2~T was used to magnetize the sample under Geometry A, while another permanent magnet with an out-of-plane field of 0.4~T was utilized to magnetize the sample under Geometry B. A Scienta Omicron DA30L spectrometer was used to analyze the emitted photoelectrons. During the spin-resolved ARPES measurements, VLEED (very low-energy electron diffraction) detectors were utilized with the spin quantization axes selectively probing the out-of-plane $P_{[110]}$ and the in-plane $P_{[001]}$ and $P_{[1\bar{1}0]}$ directions. The film was aligned to have negligible angle offsets relative to the analyzer ($\Gamma$ at $k_x=k_y=0$) when measured under deflector mode. The spin polarization is calculated from 
\begin{equation}
    P = \frac{1}{S}\frac{I_\uparrow-I_\downarrow}{I_\uparrow+I_\downarrow},
    \label{eq:sarpes_P}
\end{equation}
where $S$ is the Sherman function taking the value of 0.2 during the time of the measurements. The corresponding spin-up and spin-down EDCs were measured up to the same acquisition time and normalized by the area using the counts within [55.3, 55.8] eV kinetic energies for data taken under 62 eV light, and within [48.3, 48.8] eV kinetic energies for data taken with 55 eV photon energy. These kinetic energy ranges have been observed dominated by the photoemission background signals in the spin-integrated mode, therefore suitable for the normalization of the spin-resolved EDCs. The error bars of the spin polarization are calculated using the error propagation formula: 
\begin{equation}
    \delta P = P \cdot \sqrt{\frac{(\sqrt{I_\uparrow})^2+(\sqrt{I_\downarrow})^2}{(I_\uparrow+I_\downarrow)^2}+\frac{(\sqrt{I_\uparrow})^2+(\sqrt{I_\downarrow})^2}{(I_\uparrow-I_\downarrow)^2}},
    \label{eq:sarpes_err}
\end{equation}
where the uncertainty of the spin-resolved photoelectron counts takes the form of $\sqrt{I_\uparrow}$ and $\sqrt{I_\downarrow}$ assuming the Poisson statistics of $I_\uparrow$ and $I_\downarrow$.

In situ x-ray photoelectron spectroscopy (XPS) was carried out inside the spin-resolved ARPES chamber. A pass energy of 10~eV was used to collect the XPS spectra within the kinetic energy range of 20 to 55~eV under 325 and 320~eV photons, where peaks relevant with Ru 3$d$ and C 1$s$ can be covered. Ex situ core-levels X-ray photoelectron spectroscopy (XPS, Physical Electronics VersaProbe III) were measured with a monochromatic Al K$\alpha$ X-ray source (1486.6 eV), where a flood gun was used to prevent photoemission-induced surface charge effects.

The ARPES data of the 14nm RuO2 on TiO2 substrate were collected at the ULTRA endstation at the Surface/Interface Spectroscopy (SIS) beamline of the Swiss Light Source, Paul Scherrer Institute. The data were acquired with a Scienta Omicron DA30L hemispherical analyzer. The energy and angular resolution are better than 20~meV and 0.1$^\circ$. The measurements were performed at a temperature of 20~K in a base pressure better than 1$\times$10$^{-10}$~Torr. The as-received RuO$_2$ films were post-annealed at 560~$^\circ$C with oxygen pressure 1$\times$10$^{-5}$ mbar for 30 minutes before ARPES measurement.

\subsubsection*{Ambient pressure X-ray photoelectron spectroscopy}

The AP-XPS measurement was carried out using a synchrotron-based AP-XPS system at the IOS (23-ID-2) beamline of the National Synchrotron Light Source II, Brookhaven National Laboratory (NSLS-II, BNL, United States), which consisted of a differentially pumped electrostatic lens and a hemispherical electron analyzer (Specs GmbH, Phoibos 150 NAP). A focused monochromatic X-ray beam with dimensions of 80 $\mu m$ (horizontal) × 20 $\mu m$ (vertical) and a resolving power (E/$\delta$E) of 10000 was employed \cite{Waluyo2022}.  The sample was annealed under oxygen conditions at 600 K for 30 minutes to confirm oxygen annealing effect of RuO$_2$ surface bonding states. The base pressure in the analysis chamber was $1\times10^{-9}$ Torr, and the oxygen pressure was maintained at 100 mTorr during the annealing process. A pyrolytic boron nitride heater was used for annealing, and a K-type thermocouple was attached to the sample surface for temperature monitoring.

\subsubsection*{First-principles calculations}

The spin-resolved electronic structure of bulk RuO$_2$ was calculated by first-principles calculations based on density functional theory~\cite{Kohn1965} as implemented in Vienna \textit{ab initio} simulation package (VASP)~\cite{Kresse1996}. 
We employed the projector-augmented wave pseudopotentials~\cite{Blochl1994,Kresse1999} and the generalized gradient approximation of Perdew-Burke-Ernzerhof (PBE)~\cite{Perdew1996} exchange-correlation (XC) functional. The kinetic energy cutoff for the plane wave basis was chosen to be 500~eV and the Brillouin zone was sampled by 20$\times$20$\times$16 $k$-mesh grid for the primitive unit cell of bulk RuO$_2$. To intuitively compare the experimental ARPES results of the RuO$_2$ thin film with the theoretical bulk band structure, we plot the band structure of bulk RuO$_2$ at a fixed $k_z=2\pi/6d$. This specific $k_z$ value was selected to effectively describe the quantum confinement effects in the RuO$_2$ thin film, using a discretized $k_z$ sampling method~\cite{kawasaki2018engineering}.

The inversion-symmetric slab calculations of the strained 15-layer RuO$_2$ and 29-layer RuO$_2$/TiO$_2$ were carried out using the full-potential (linearized) augmented-plane-wave plus local orbitals implementation of density functional theory~\cite{Kohn1965}, WIEN2k~\cite{Blaha2020}. The experimental crystal structure of bulk TiO$_2$ were used to construct both the slabs. A vacuum region of 26.2$\textup{~\AA}$ and 33.9$\textup{~\AA}$ were added into the 15-layer RuO$_2$ and 29-layer RuO$_2$/TiO$_2$ slabs, respectively. The respective slabs have 90 (39 independent) and and 174 (74 independent) atoms in the unit cell. All the presented calculations were non-spin-polarized. The structural relaxation was carried out using the generalized gradient approximation~\cite{Perdew1996} for the exchange-correlation functional. All the atomic positions not fixed by the $Pmmm$ space group symmetries were allowed to relax. Therefore, out-of-plane partial strain relaxation could be captured in the calculations. Muffin-tin radii of 1.87~$a_0$, 1.83~$a_0$, and 1.61~$a_0$ were chosen for the Ru, Ti, and O atoms, respectively, $a_0$ being the Bohr radius. The plane-wave cutoff criteria were set to $G_{max}$~=~16, $R_{MT}K_{max}$~=~6.50 for the 15-layer strained RuO$_2$, and $R_{MT}K_{max}$~=~6.23 for the 29-layer RuO$_2$/TiO$_2$. The non-spherical matrix elements were expanded up to $l$ = 6. A $k$-mesh of $16\times7\times1$ was adopted in the full Brillouin zone. The Fermi level (\ef) was calculated under a Fermi function broadening with a broadening parameter of 0.002 Ry. A simultaneous convergence of energy, charge, and force was required for the convergence of the self-consistent field (SCF) calculations. For the 15-layer strained RuO$_2$, it was set to $10^{-4}$~Ry, $10^{-3}~e^-$, and 0.5~mRy~a.u.$^{-1}$. For the 29-layer RuO$_2$/TiO$_2$, it was set to $10^{-4}$~Ry, $5\times10^{-3}~e^-$, and 1.0~mRy~a.u.$^{-1}$. The charge convergence is only reached when three consecutive SCF iterations are below the charge threshold. A $k$-mesh convergence test was performed for the non-spin-polarized strained 15-layer RuO$_2$ using a grid of $32\times14\times1$ with a Fermi broadening of 0.0018~Ry where a charge convergence of $5\times10^{-4}~e^-$ was achieved. Nonetheless, the relaxed crystal structure and band structure were affected negligibly. The slab calculations employ scalar-relativistic approximation, while the bulk calculations (purple markers in Fig.~\ref{fig:fig3} (C and E)) include spin-orbit coupling. Due to the truncated (110) surface Brillouin zone in the slab geometry and without the focus on band unfolding spectral weight, slab band structures along $\Bar{X}-\Bar{M}$ and $\Bar{Y}-\Bar{A}$ in main text Fig.~\ref{fig:fig3} are mirror reflected from $\Bar{\Gamma}-\Bar{X}$ and $\Bar{Z}-\Bar{Y}$, respectively. 

The one-step model ARPES calculations were performed using the spin-polarized relativistic Korringa-Kohn-Rostoker (SPRKKR) package~\cite{Ebert2011}, implementing the option of fully relativistic four component Dirac formalism under the atomic sphere approximation. The generalized gradient approximation of PBE~\cite{Perdew1996} was used for the exchange-correlational functional and the crystal structure adopted a WIEN2k relaxed bulk RuO$_2$ structure under the strain provided by experimental TiO$_2$ lattice constants. We used 32 energy points on the Gaussian-Legendre quadrature path for the energy integration and a $k$-mesh of $17\times17\times27$ for the Brillouin zone integration during the self-consistent field (SCF) calculations. The angular momentum summation was cut off at $l_{max}=4$. The Lloyd's formula was used for determining the Fermi level. Since the goal is to examine the mirror parities of the photoelectron spin polarization under a paramagnetic assumption in Table.~\ref{tab:sarpes}, we imposed a nonmagnetic constraint during the SCF cycles and no Hubbard $U$ correction was used. For the photoemission calculations, the strained bulk crystal was terminated on the stoichiometric (110) surface, employing a photoemission Geometry B specified in fig.~\ref{fig:fig_geometry_B}. The final states were modeled as time-reversed low-energy electron diffraction states and the spin-polarized photoemission intensity was calculated under the spin-density matrix formalism~\cite{Braun2018}. Finite lifetime of the initial and final states was simulated with an imaginary potential of 0.01 and 1.5~eV, respectively. The surface potential used a Rundgren-Malmström type $z$-dependent potential~\cite{J.RundgrenandG.Malmstrom1977} and both cases of considering and disregarding surface reflections were tested yielding the same mirror-parity conclusions shown in fig.~\ref{fig:fig_one_step}.


\subsection*{Supplementary Materials}
\textbf{This PDF file includes:}\\
Supplementary Text\\
Figs. S1 to S19\\
Table S1\\
References\\
Cif file of the strained RuO$_2$ slab\\
Cif file of the strained RuO$_2$/TiO$_2$ slab




%
\bibliography{main} 
\bibliographystyle{sciencemag}

%
%
%
%
%
%


\section*{Acknowledgments}
The authors thank Aki Pulkkinen, Alberto Marmodoro, Jonathan A. Sobota, Dongyu Liu, Ji Seop Oh and Byong Ki Choi for helpful discussions.
\paragraph*{Funding:}
The ARPES work at Rice University was supported by the U.S. Department of Energy, BES grant No. DE-SC0026179, the Gordon and Betty Moore Foundation’s EPiQS Initiative through grant No. GBMF9470 and the Robert A. Welch Foundation Grant No. C-2175 (M.Y.). Film synthesis (S.G.J and B.J.) was supported by the U.S. Department of Energy through grant Nos. DE-SC0020211, and (partly) DE-SC0024710. 
S.N. was supported partially by the UMN MRSEC program under Award No. DMR-2011401. Parts of this work were carried out at the Characterization Facility, University of Minnesota, which receives partial support from the NSF through the MRSEC program under Award No. DMR-2011401. B.J. and S.G.J. also thank partial support by the Air Force Office of Scientific Research (AFOSR) through Grant Nos. FA9550-21-1-0025, and FA9550-24-1-0169. Y.Z. acknowledges support from the US National Science Foundation (NSF) Grant Number 2201516 under the Accelnet program of Office of International Science and Engineering (OISE). This research used resources of the Advanced Light Source, which is a DOE Office of Science User Facility under contract no. DE-AC02-05CH11231. R.M.F. was supported by the Air Force Office of Scientific Research under Award No. FA9550-21-1-0423. Ab-initio calculations at Rice were supported in part by the NOTS cluster operated by Rice University's Center for Research Computing (CRC). M.R. acknowledges support from the Swiss National Science Foundation - SNSF (Project No. 10001679). J.M. thanks the project Quantum Materials for applications in sustainable technologies (QM4ST), funded as Project No. CZ.02.01.01/00/22\_008/0004572 by Programme Johannes Amos Comenius, call Excellent Research. I.H.C. and J.S.L were supported by the National Research Foundation of Korea (NRF) grant funded by the Korean government (MSIT) (grant no. RS-2024-00486846). The theory work at the University of Minnesota (L.B. and T.B.) was supported by the NSF CAREER grant DMR-2046020. AP-XPS work (H.L., I.W., and A.H.) used resources of the 23-ID-2 (IOS) beamline of the National Synchrotron Light Source II, a U.S. Department of Energy (DOE) Office of Science User Facility operated for the DOE Office of Science by Brookhaven National Laboratory under Contract No. DE-SC0012704.

\paragraph*{Author contributions:}
B.J., M.R., and M.Y. oversaw the project. Y.Z. performed the ARPES and in situ XPS experiments on the 2~nm films, with help from J.W., Y.G., Z.R., Z.Y., J.H., and T.Z., supervised by J.K. and M.Y., and with beamline support from A.F. and S.K.M.. H.Li and M.R. performed the ARPES data acquisition and analysis on the 14~nm films. Y.Z. performed the spin-resolved ARPES geometry analyses, in discussion with M.Y. and J.M.. Y.Z. and J.M. performed the one-step model ARPES calculations. Y.Z. performed the inversion-symmetric slab DFT calculations.  S.L. and T.L. performed the bulk DFT calculations. L.B., T.B. and R.M.F. carried out the group theoretical analysis of the experimental results. S.G.J., S.N., and B.J. designed and synthesized the epitaxial oxide heterostructures using hMBE. S.G.J. and B.J. performed RHEED, XRD, and AFM for characterizing the film quality. S.G.J. and B.J. performed core-level XPS measurements and data analysis. I.H.C. and J.S.L. conducted optical SHG measurements and data analysis. H.Lim, I.W., and A.H. conducted AP-XPS measurements. All authors contributed to the manuscript writing.
\paragraph*{Competing interests:}
The authors declare that they have no competing interests.
\paragraph*{Data, code, and materials availability:}
All data and code needed to evaluate and reproduce the results in the paper are present in the paper and/or the Supplementary Materials. This study did not generate new materials.


\begin{figure}
    \includegraphics[width=0.7\textwidth]{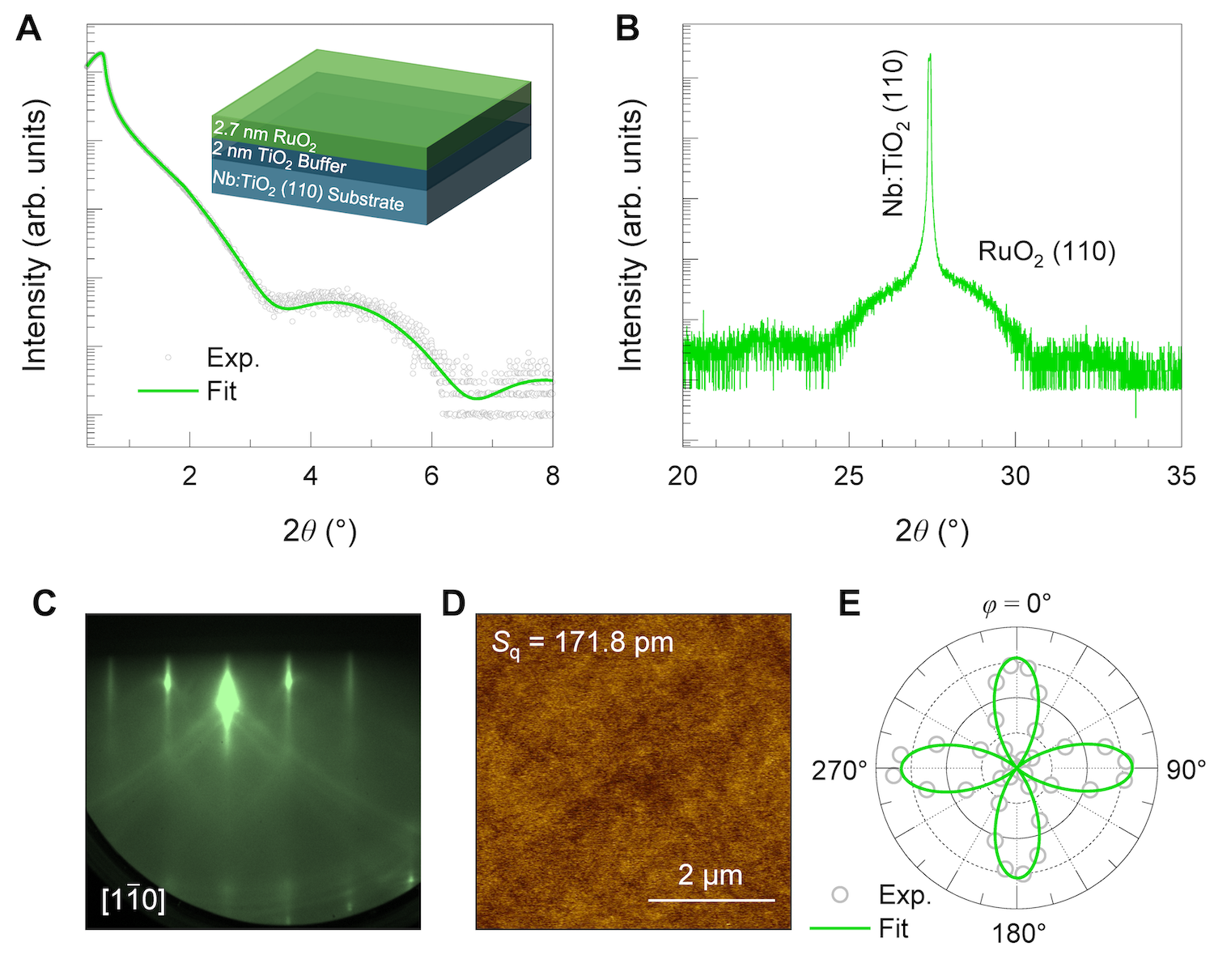}
    \centering
    \caption{\textbf{Design and structural characterization of fully strained metallic RuO$_2$ (110) heterostructures grown by hMBE.} (\textbf{A}) XRR and (\textbf{B}) XRD 2$\theta$-$\theta$ scans of RuO$_2$ heterostructures. Scattered symbols and solid lines in (A) represent the experimental data and corresponding fitting results, respectively. The inset in (A) shows a schematic illustration of the heterostructure architecture comprising of 2.7 nm RuO$_2$/ 2 nm TiO$_2$/ Nb:TiO$_2$ (110). (\textbf{C}) RHEED patterns acquired after growth along the [1$\bar{1}$0] direction reveal streaky features with Kikuchi lines, indicative of high crystalline quality. (\textbf{D}) AFM image demonstrating atomically smooth surface morphology. (\textbf{E}) Representative rotational anisotropy SHG results with both fundamental and SHG polarizations parallel to the incidence plane. Fitting curves (solid lines) based on noncentrosymmetric \textit{mm}2 symmetry agree well with the experimental data (scattered symbols). Full polarization analysis is included in the Supplementary Text.}
    \label{fig:fig2}
\end{figure}

\begin{figure}
    \includegraphics[width=0.6\textwidth]{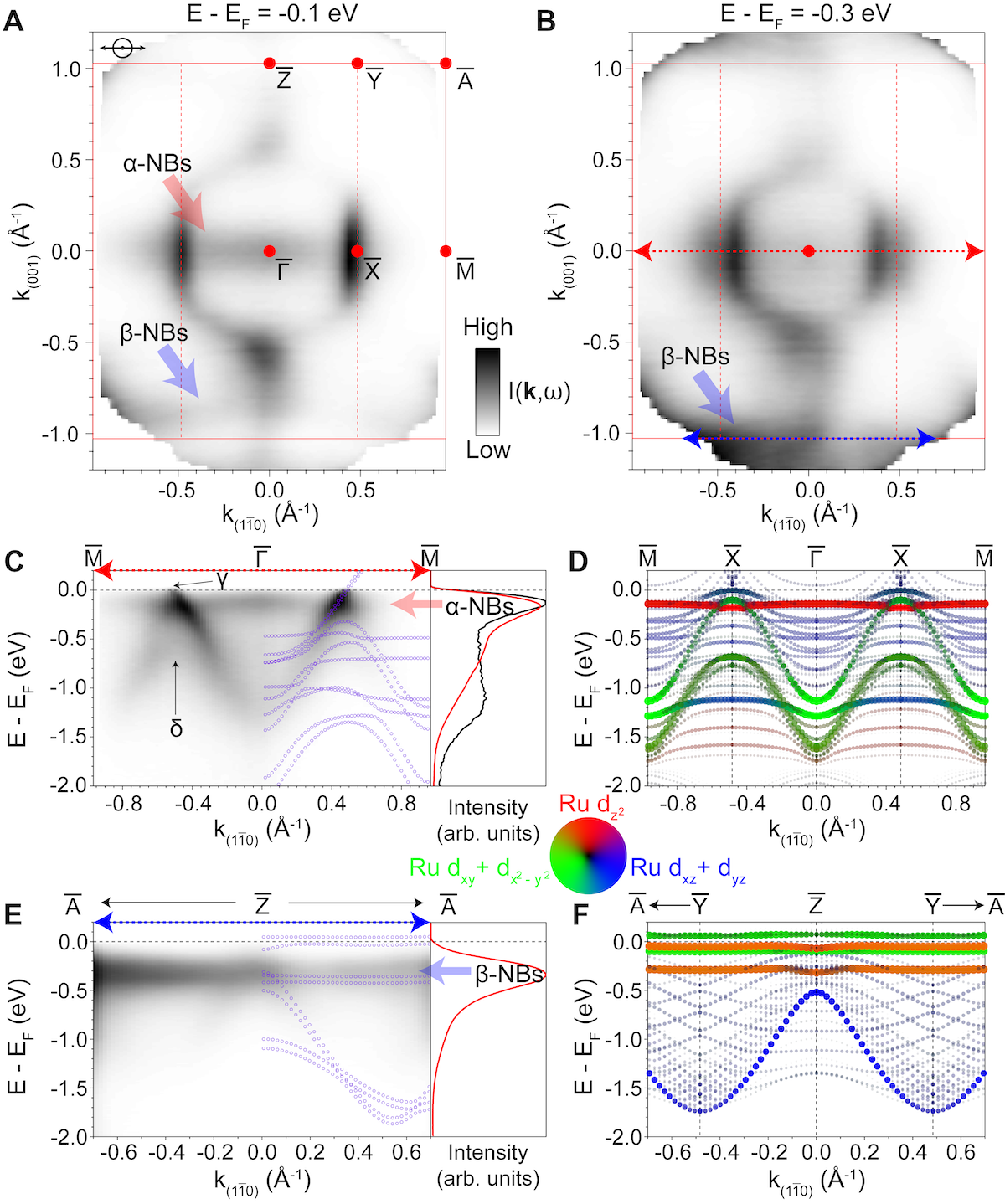}
    \centering
    \caption{\textbf{Narrow bands (NBs) in the ultrathin epitaxially strained RuO$_2$.} (\textbf{A}) Constant energy contour (CEC) at $E$~-~\ef~=~-0.1~eV measured by 62 eV $p$-polarized (indicated on the left top) photons, with an energy integration window of 20~meV, emphasizing the $\alpha$ narrow bands ($\alpha$-NB) and $\beta$ narrow bands ($\beta$-NB) denoted by the red and blue arrows. (\textbf{B}) CEC at $E$~-~\ef~=~-0.3~eV showing the $\beta$-NB. (\textbf{C}) Measured electronic band dispersions along the high symmetry $\Bar{\Gamma}-\Bar{M}$ direction spanned by the red dotted arrow in (B), overlaid with spin-polarized bulk density functional theory (DFT) calculations carried out with full TiO$_2$ substrate strain. The energy distribution curve (EDC) integrated across the presented momentum range is shown on the right in red, while the black EDC is the single EDC at $\Bar{\Gamma}$. (\textbf{D}) Non-spin-polarized DFT calculated $\Bar{\Gamma}-\Bar{M}$ band structure based on a 15-layer inversion-symmetric fully strained RuO$_2$ slab (see Supplementary Text for more details). The meaning of the size and transparency of the markers have been specified in the main text. The red, green, and blue colors indicate the projection weights onto the respective $d_{z^2}$, $d_{xy}+d_{x^2-y^2}$, and $d_{xz}+d_{yz}$ orbitals of the Ru atoms within the two surface layers. (\textbf{E} and \textbf{F}) Same as (C and D) but}
    \label{fig:fig3}
\end{figure}

\clearpage

\vspace*{0pt}
\noindent{\textbf{(continued caption)} along the high symmetry $\Bar{Z}-\Bar{A}$ direction spanned by the blue dotted arrow in panel (B). All measurements in the main text, unless otherwise specified, were performed at 15 K and with the same $p$-polarized light. All $k_x$-$k_y$ maps in the main text integrate 20~meV in energy.}

\begin{figure}
    \includegraphics[width=\textwidth]{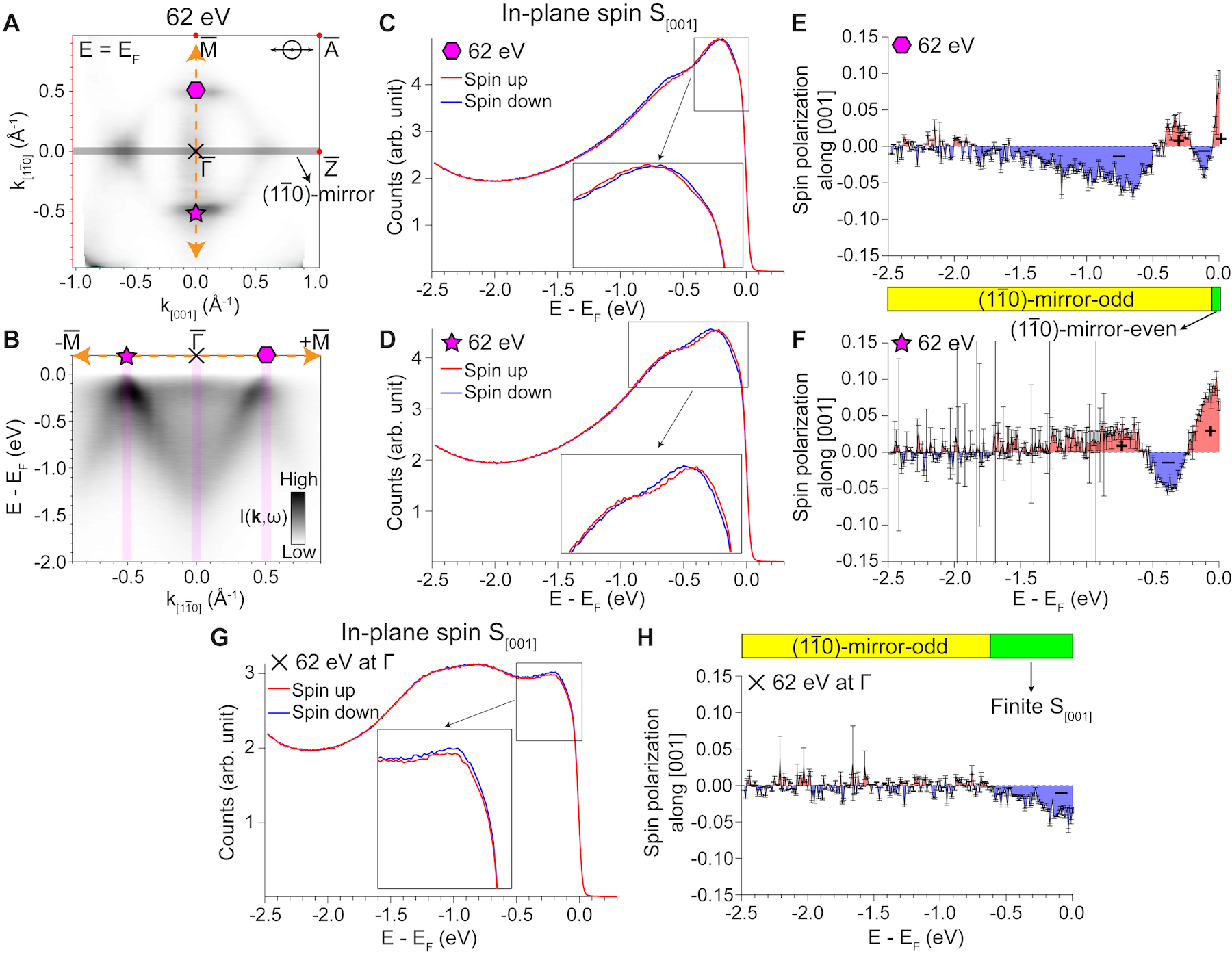}
    \centering
    \caption{\textbf{Measured photoelectron spin polarization along the in-plane [001] direction on the $\Bar{\Gamma}\Bar{M}$ path under Geometry A.} (\textbf{A}) Fermi surface probed by 62~eV $p$-polarized photons (symbol on the top right of the panel). (\textbf{B}) Band dispersions along the $\bar{\Gamma}-\bar{M}$ direction indicated by the vertical dashed double-arrow in (A). The magenta and the cross symbols in (A and B) indicate where the spin-resolved energy distribution curves (EDCs) are taken. (\textbf{C}, \textbf{D}, \textbf{G}) Raw spin-resolved EDCs taken on both sides of the (1$\bar{1}$0)-mirror, as well as at $\Bar{\Gamma}$, selecting photoelectrons with spin polarization only along the in-plane [001] direction. (\textbf{E}, \textbf{F}, \textbf{H}) Converted spin polarization based on (C, D, G), respectively. See Materials and Methods for details about the background normalization and error bars.}
    \label{fig:fig4}
\end{figure}

\clearpage

\begin{figure}
    \includegraphics[width=\textwidth]{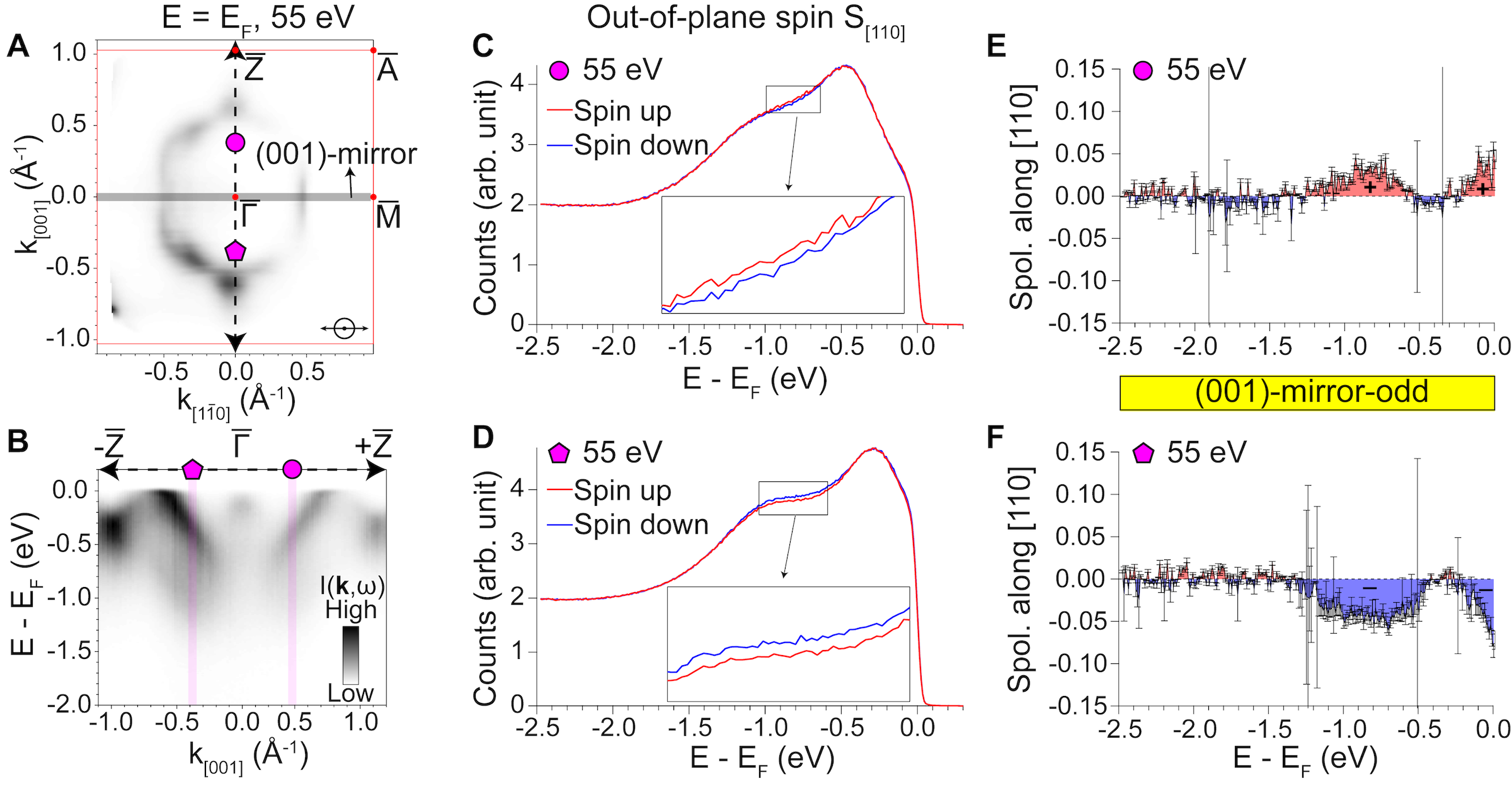}
    \centering
    \caption{\textbf{Out-of-plane photoelectron spin polarization on the $\Bar{\Gamma}\Bar{Z}$ path perpendicular to the (001)-mirror plane measured in Geometry B.} (\textbf{A}) Fermi surface under 55~eV photons highlighting the (001)-mirror and the momentum positions of the measured spin-resolved energy distribution curves (EDCs) using circular and pentagonal symbols. (\textbf{B}) Electronic band dispersions measured along $\Bar{\Gamma}-\Bar{Z}$. Similarly, the width of the vertical magenta bars provides a visualization of the momentum resolution in the spin-resolved measurement mode. (\textbf{C} and \textbf{D}) Raw spin-resolved EDCs selectively probing only the out-of-plane [110] spin polarization on the upper and lower sides of the (001)-mirror. (\textbf{E} and \textbf{F}) Converted out-of-plane spin polarization from (C and D). See Materials and Methods for details about the background normalization and error bar calculation.}
    \label{fig:fig5}
\end{figure}

\clearpage

\begin{table}
    \centering
    \caption{\textbf{Extrinsic final state selection rules for spin-resolved ARPES for our measurement geometry in the nonmagnetic \textit{mm2} phase of RuO$_2$.} The light incidence considers $p$-polarized photons aligned within a mirror plane of the crystal off-normally. The right three columns specify for the nonmagnetic \textit{mm2} phase of RuO$_2$ whether the three orthogonal components of the final state photoelectron spin polarization ($P$) are allowed to exhibit a finite mirror-even component after being reflected by the preserved mirror plane of the total photoemission system (light + semi-infinite crystals + photoelectrons).}
    \label{tab:sarpes}
    
    \begin{tabular}{lcccccc}
        \\
        \hline
        Name&Light incidence&$m_{(1\Bar{1}0)}$&$m_{(001)}$&$P_{[001]}$&$P_{[1\Bar{1}0]}$&$P_{[110]}$\\
        \hline
        Geometry A&$m_{(1\Bar{1}0)}$&preserved&broken&not allowed&allowed&not allowed\\
        Geometry B&$m_{(001)}$&broken&preserved&allowed&not allowed&not allowed\\
        \hline
    \end{tabular}
\end{table}
\clearpage

\begin{table} 
	\centering
        \footnotesize
	\caption{\textbf{Classification of constant, even-in-$k$, and odd-in-$k$ spin-splitting terms as irreducible representations of the paramagnetic point group $\textit{mm}$2.1' up to second order in $\textit{k}$.}} Terms dressed with underlines (bold fonts) are those clearly observed in our experiments performed in Geometry A (B). Terms with both underlines and bold fonts were observed in both geometries.
	\label{tab:irreps} 
	
	\begin{tabular}{lccc} 
		\\
		\hline
		$mm2.1'$&$\sigma_i$&$k_i\sigma_j$&$k_ik_j\sigma_h$\\
		\hline
		$A_1^-(mm2.1)$ & $\cdot$ & $\cdot$ & $k_{110} k_{1\bar{1}0}\sigma_{001},\, k_{110}k_{001}\sigma_{1\bar{1}0},\, k_{1\bar{1}0}k_{001}\sigma_{110}$\\
		$A_2^-(m'm'2)$ & $\sigma_{110}$ & $\cdot$ & $k_{110}k_{001}\sigma_{001},\, k_{110}k_{1\bar{1}0}\sigma_{1\bar{1}0},\, (k^2_{1\bar{1}0}-k^2_{001})\sigma_{110},\,       k^2_{110}\sigma_{110}$\\
		$B_1^-(m'm2')$ & $\mathbf{\underline{\sigma_{001}}}$ & $\cdot$ & $k_{110} k_{001}\sigma_{110},\, k_{1\bar{1}0}k_{001}\sigma_{1\bar{1}0},\, \mathbf{\underline{{(k^2_{1\bar{1}0}-k^2_{001})\sigma_{001}}}},\, k^2_{110}\sigma_{001}$\\
		$B_2^-(mm'2')$ & $\mathbf{{\sigma_{1\bar{1}0}}}$ & $\cdot$ & $k_{110}k_{1\bar{1}0}\sigma_{110},\, k_{1\bar{1}0}k_{001}\sigma_{001},\, \mathbf{{(k^2_{1\bar{1}0}-k^2_{001})\sigma_{1\bar{1}0}}},\, k^2_{110}\sigma_{1\bar{1}0}$\\
		$A_1^+(mm2.1')$ & $\cdot$ & $\underline{{k_{1\bar{1}0}\sigma_{001}}},\, \mathbf{{k_{001}\sigma_{1\bar{1}0}}}$ & $\cdot$\\
		$A_2^+(2.1')$ & $\cdot$ & $k_{001}\sigma_{001},\, k_{1\bar{1}0}\sigma_{1\bar{1}0},\, k_{110}\sigma_{110}$ & $\cdot$\\
		$B_1^+(m.1')$ & $\cdot$ & $\mathbf{{k_{001}\sigma_{110}}},\, k_{110}\sigma_{001}$ & $\cdot$\\
		$B_2^+(m.1')$ & $\cdot$ & $k_{1\bar{1}0}\sigma_{110},\, k_{110}\sigma_{1\bar{1}0}$ & $\cdot$\\
		\hline
	\end{tabular}
\end{table}

\clearpage


\newpage


\renewcommand{\thefigure}{S\arabic{figure}}
\renewcommand{\thetable}{S\arabic{table}}
\renewcommand{\theequation}{S\arabic{equation}}
\renewcommand{\thepage}{S\arabic{page}}
\setcounter{figure}{0}
\setcounter{table}{0}
\setcounter{equation}{0}
\setcounter{page}{1} 


\begin{center}
\section*{Supplementary Materials for\\ \scititle}

Yichen Zhang$^{\dagger}$, Seung Gyo Jeong$^\dagger$, Luca Buiarelli, Seungjun Lee, Yucheng Guo,\\
Jiaqin Wen, Hang Li, Sreejith Nair, In Hyeok Choi, Zheng Ren, Ziqin Yue, \\
Jounghoon Hyun, Tieqiong Zhang, Alexei Fedorov, Sung-Kwan Mo, Hojoon Lim, \\
Adrian Hunt, Iradwikanari Waluyo, Junichiro Kono, Ján Minár, Jong Seok Lee,\\
Tony Low, Turan Birol, Rafael M. Fernandes, Milan Radovic$^{\ast}$,\\
Bharat Jalan$^{\ast}$, and Ming Yi$^{\ast}$\\
\small$^{\ast}$Corresponding emails:  milan.radovic@psi.ch, bjalan@umn.edu, mingyi@rice.edu\\
\small$^\dagger$These authors contributed equally to this work.
\end{center}

\subsubsection*{This PDF file includes:}
Supplementary Text\\
Figures S1 to S19\\
Table S1\\
References

\subsubsection*{Other Supplementary Materials for this manuscript:}
File S1\\
File S2

\newpage

\newpage
\subsection*{Supplementary Text}

\subsubsection*{Additional film characterization}

\begin{figure}
    \centering
    \includegraphics[width=0.55\linewidth]{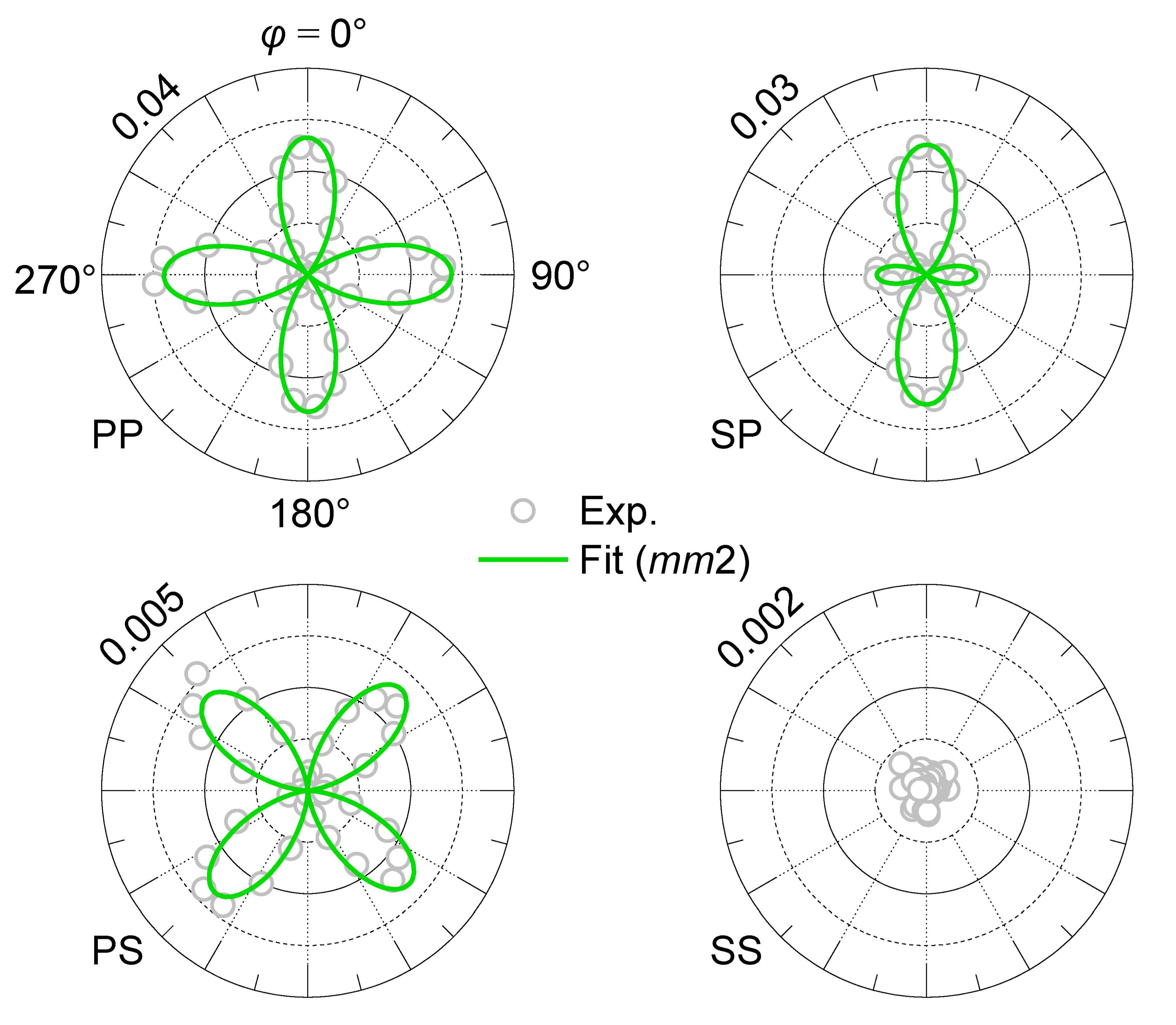}
    \caption{\textbf{Rotational anisotropy second-harmonic generation (SHG) results for RuO$_2$ heterostructures.} SHG intensity was measured as a function of azimuthal angle $\phi$ under an oblique incidence of 45$^\circ$ at room temperature. The incident fundamental light was set to either P- or S-polarization (P$_{\rm in}$ or S$_{\rm in}$), and the SHG signal was detected in both P- and S-polarizations (P$_{\rm out}$ or S$_{\rm out}$), where P and S denote polarization parallel and perpendicular to the plane of incidence, respectively. The fitting curves (solid lines), based on a noncentrosymmetric $mm2$ point group with electric-dipole allowed SHG, show excellent agreement with the experimental data (scattered symbols). A detailed symmetry analysis can be found in our previous study~\cite{Jeong2025}.}
    \label{fig:fig_SHG_more}
\end{figure}

Figure~\ref{fig:fig_SHG_more} presents rotational anisotropy second-harmonic generation (RA-SHG) measurements, consistent with the  $mm2$ ($C_{2v}$) point group symmetry, indicating a fully strained, noncentrosymmetric RuO$_2$ heterostructure. The RA-SHG patterns in the PP, SP, and PS configurations exhibit a two-fold rotational axis along the [110] direction, with two mirror planes normal to the [1$\bar{1}$0] and [001] directions. The absence of SHG response in the SS configuration, combined with these symmetry elements, confirms that the SHG signal originates from the electric dipole contribution under the noncentrosymmetric $mm2$ point group, where the breaking of the mirror symmetry perpendicular to the [110] axis.

\begin{figure}
    \centering
    \includegraphics[width=0.45\linewidth]{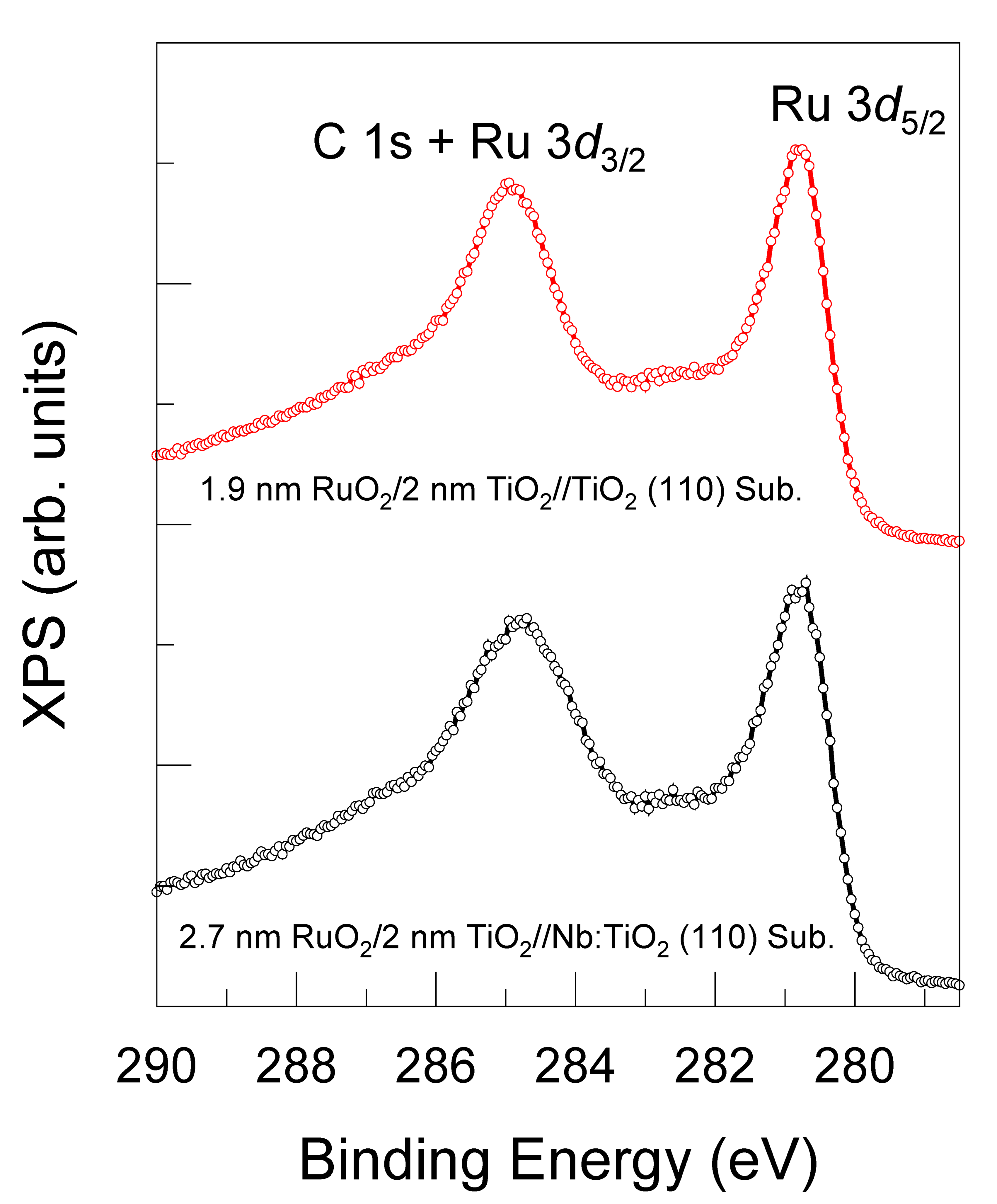}
    \caption{\textbf{Ru 3$d$ x-ray photoelectron spectroscopy (XPS) spectra of RuO$_2$ heterostructures grown on TiO$_2$ and Nb:TiO$_2$ substrates.} Although the Ru 3$d_{3/2}$ peak overlaps with the C 1$s$ peak due to their close binding energies, the consistent Ru 3$d_{5/2}$ peak position and width in both samples confirm the consistent oxidation state. These are also consistent our previous XPS study of RuO$_2$ samples~\cite{William_2021}.}
    \label{fig:fig_XPS}
\end{figure}

\begin{figure}
    \centering
    \includegraphics[width=0.9\linewidth]{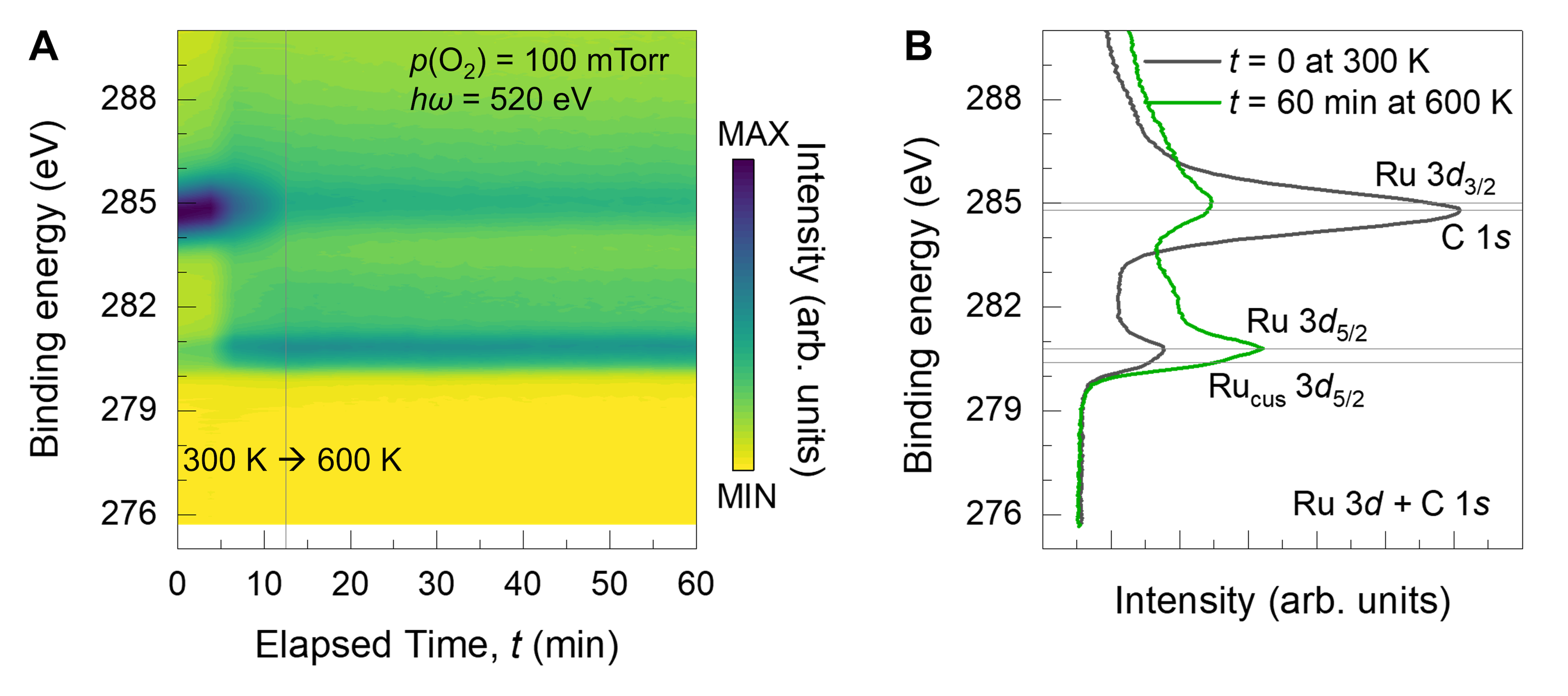}
    \caption{{\textbf{Oxygen annealing study of RuO\textsubscript{2}/TiO\textsubscript{2} (110) epitaxial thin films with AP-XPS.} }(\textbf{A}) Contour plots of APXPS spectra at Ru 3\textit{d} and C 1\textit{s} core levels as a function of elapsed time (\textit{t}) and binding energy during oxygen annealing process. We have shipped samples to APXPS facilities and loaded UHV chamber. Subsequently, we have annealed sample at 600 K and 100 mTorr of \textit{p}(O\textsubscript{2}) in APXPS chamber; the temperature reached 600 K at \textit{t} = 13 min, approximately. Acquisition time of each spectrum is 43 seconds, and photon energy ($h\omega$) is used to 520 eV. (\textbf{B}) APXPS spectra at \textit{t} = 0 and 60 minutes. Upon oxygen annealing, the C 1\textit{s} intensity is strongly suppressed, while the Ru 3$d$ intensity increases, indicating effective surface cleaning. Notably, the binding energy position of the Ru 3$d_{5/2}$ peak remains identical, demonstrating that the Ru-O chemical bonding is well preserved during annealing. In addition, a shoulder feature corresponding to coordinatively unsaturated Ru atoms (Ru\textsubscript{cus} 3$d_{5/2}$), previously reported in XPS studies~\cite{OVER2001}, is observed, suggesting the presence of Ru\textsubscript{cus} sites at the surface. }
    \label{fig:fig_APXPS}
\end{figure}

To characterize the surface stoichiometry of the RuO$_2$ films, x-ray photoelectron spectroscopy (XPS) measurements were carried out under different conditions. First, as shown in Fig.~\ref{fig:fig_XPS}, ex situ Ru 3$d$ core-level XPS spectra of RuO$_2$ heterostructures grown on both TiO$_2$ and Nb:TiO$_2$ (with TiO$_2$ buffer layer) substrates consistently exhibit +4 oxidation states of Ru atoms, confirming the chemical stability across different substrates. However, the presence of carbon contamination in the ex situ environment is present, as can be seen from the relatively large peak area of the 285~eV binding energy peak relative to the one of the 280.8~eV peak in Fig.~\ref{fig:fig_XPS}. Therefore, next we explore the oxygen annealing procedures on the RuO$_2$/TiO$_2$ (110) epitaxial films using ambient-pressure-XPS. As shown in Fig.~\ref{fig:fig_APXPS} (A and B), annealing under an oxygen pressure of 100~mTorr for 5 min at 600~K strongly suppresses the spectral intensity in the binding-energy region where the Ru $3d_{3/2}$ and C $1s$ signals overlap, indicating effective removal of surface carbon contamination. As a result, the apparent intensity of the Ru $3d_{5/2}$ component becomes dominant, and the intensity ratio between the Ru $3d_{5/2}$ and Ru $3d_{3/2}$ peaks approaches the expected 3:2 spin–orbit branching ratio. This indicates the restoration of a close-to-stoichiometric RuO$_2$ (110) surface~\cite{Sasaki1999,Kaga1999,OVER2001}. 

\begin{figure}
    \centering
    \includegraphics[width=0.95\linewidth]{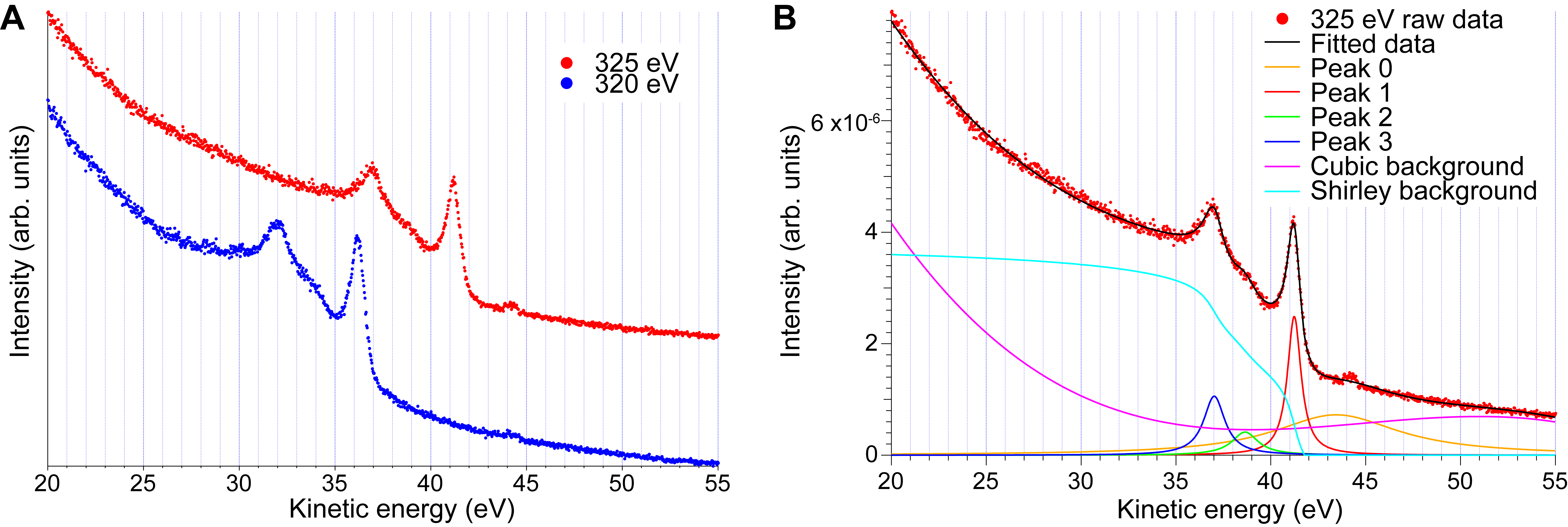}
    \caption{\textbf{In situ x-ray photoelectron spectroscopy (XPS) of the 2~nm RuO$_2$/TiO$_2$ epitaxial film measured by spin-resolved angle-resolved photoemission experiments.} (\textbf{A}) Raw XPS data targeting Ru 3$d$ and C 1$s$ binding energy ranges under 325~eV and 320~eV photons. (\textbf{B}) The fitting results of the 325~eV data involving Lorentzian peaks, Shirley background, and a polynomial background up to the third order.}
    \label{fig:fig_insitu_XPS}
\end{figure}

Finally, we perform XPS measurements inside the spin-resolved ARPES chamber for the oxygen-annealed 2~nm RuO$_2$/TiO$_2$ films. As shown in Fig.~\ref{fig:fig_insitu_XPS}A, we observe an XPS curve featuring a consistent suppression of carbon contamination and two main peaks belonging to Ru $3d_{5/2}$ and $3d_{3/2}$. We first provide a comparison between XPS data taken at 325 and 320~eV plotted as a function of photoelectron kinetic energy (Fig.~\ref{fig:fig_insitu_XPS}A), where peaks with a 5~eV shifted kinetic energy can be confirmed to be core-level peaks, while the 44~eV peak feature in the 325 eV data is strongly suppressed in the 320~eV data and stays at the same kinetic energy. This indicates its kinetic-independent nature and a potential origin from the Auger process. Next, using the 325 eV data, we use a polynomial plus Shirley background and four Lorentzian peaks to fit the curve across the whole measured energy range. We arrived at the best fit with a third-order polynomial background, rather than quadratic, especially for the background at low kinetic energy range. The red (peak 1) and blue (peak 3) peaks are attributed to the Ru $3d_{5/2}$ and $3d_{3/2}$ peaks of RuO$_2$, the green peak (peak 2) is included for the satellite peak of Ru $3d_{5/2}$~\cite{Over2002}, while the broad orange peak is used to cover a Ru$_{\rm cus}$ contribution~\cite{OVER2001} under the sharp $3d_{5/2}$ main peak, as well as the 44~eV Auger peak. The extracted peak area ratio between peak 1 and peak 3 is around 1.52, close to a theoretical peak area ratio of 3 : 2 for Ru $3d_{5/2}$ versus Ru $3d_{3/2}$. However, due to the freedom of including more parameters under fitting, such a peak area ratio should not be considered as direct evidence of a perfectly stoichiometric surface. Nonetheless, the in situ XPS results indicate a similar conclusion that the samples measured by spin-resolved ARPES after the oxygen annealing procedures have suppressed carbon contamination and a close-to-stoichiometric RuO$_2$ (110) surface.

\subsubsection*{First-principles calculations on strained RuO$_2$}

To gain insight into the character of the bands seen by ARPES, we perform first-principles calculations (main text Fig.~\ref{fig:fig3} (C and E)), for three distinct sets of parameters for bulk RuO$_2$: (1) fully strained without the $U$ correction, (2) unstrained without the $U$ correction, and (3) unstrained with a 2 eV $U$ correction for Ru $d$ orbitals. All these calculations included spin-orbit coupling. Our goal here is not to discuss the possible emergence of altermagnetism in DFT+$U$ calculations, nor the appropriate set of $U$ values that realistically describes the bulk material, but rather to compare the bands obtained in each case with our ARPES data.

\begin{figure}
    \centering
    \includegraphics[width=0.95\linewidth]{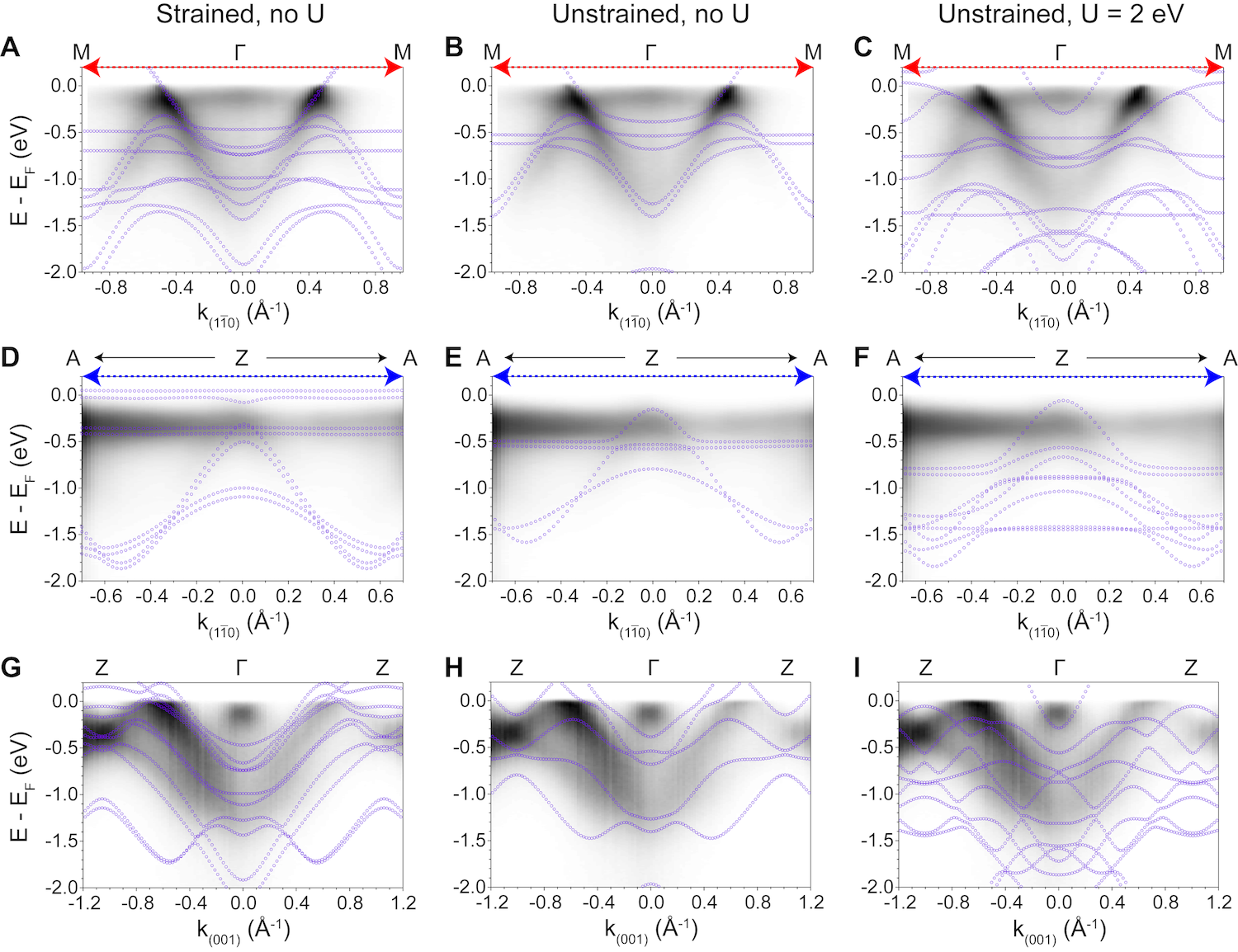}
    \caption{\textbf{Comparison of calculated strain- and Hubbard U-dependent electronic structures of RuO$_2$ with experimental ARPES spectra} (\textbf{A}) Calculated electronic structures of bulk RuO$_2$ (purple dashed line) and ARPES spectra of the 2nm RuO$_2$ thin film along (\textbf{A}-\textbf{C}) M-$\Gamma$-M, (\textbf{D}-\textbf{F}) A-Z-A, (\textbf{G}-\textbf{I}) Z-$\Gamma$-Z high-symmetry lines. The bulk RuO$_2$ electronic structures were obtained using three distinct approaches: (\textbf{A}, \textbf{D}, \textbf{G}) fully strained without Hubbard U correction, (\textbf{B}, \textbf{E}, \textbf{H}) unstrained without Hubbard U correction, and (\textbf{C}, \textbf{F}, \textbf{I}) unstrained with a 2 eV Hubbard U correction applied to the Ru d orbital. For each case, the self-consistent field ground state is denoted at the top.}
    \label{fig:fig_bulk_DFT}
\end{figure}

As shown in Fig.~\ref{fig:fig_bulk_DFT}, the set of parameters (3) exhibits the poorest agreement with experimental ARPES (Fig.~\ref{fig:fig_bulk_DFT} (C, F, I)), raising doubts on this parameter choice. While the results from the set (2) show slightly better agreement, they still demonstrate limitations, particularly about the narrow band positions (Fig.~\ref{fig:fig_bulk_DFT}E).

As shown in Fig.~\ref{fig:fig_bulk_DFT}(A, D, G), the set of parameters (1), corresponding to strained sample without a $U$ correction, shows the best agreement with the experimental results compared to the two previous cases. Consequently, our results suggest that strain plays a crucial role in determining the magnetic and electronic properties of our thin films of RuO$_2$.

\begin{figure}
    \centering
    \includegraphics[width=1.0\linewidth]{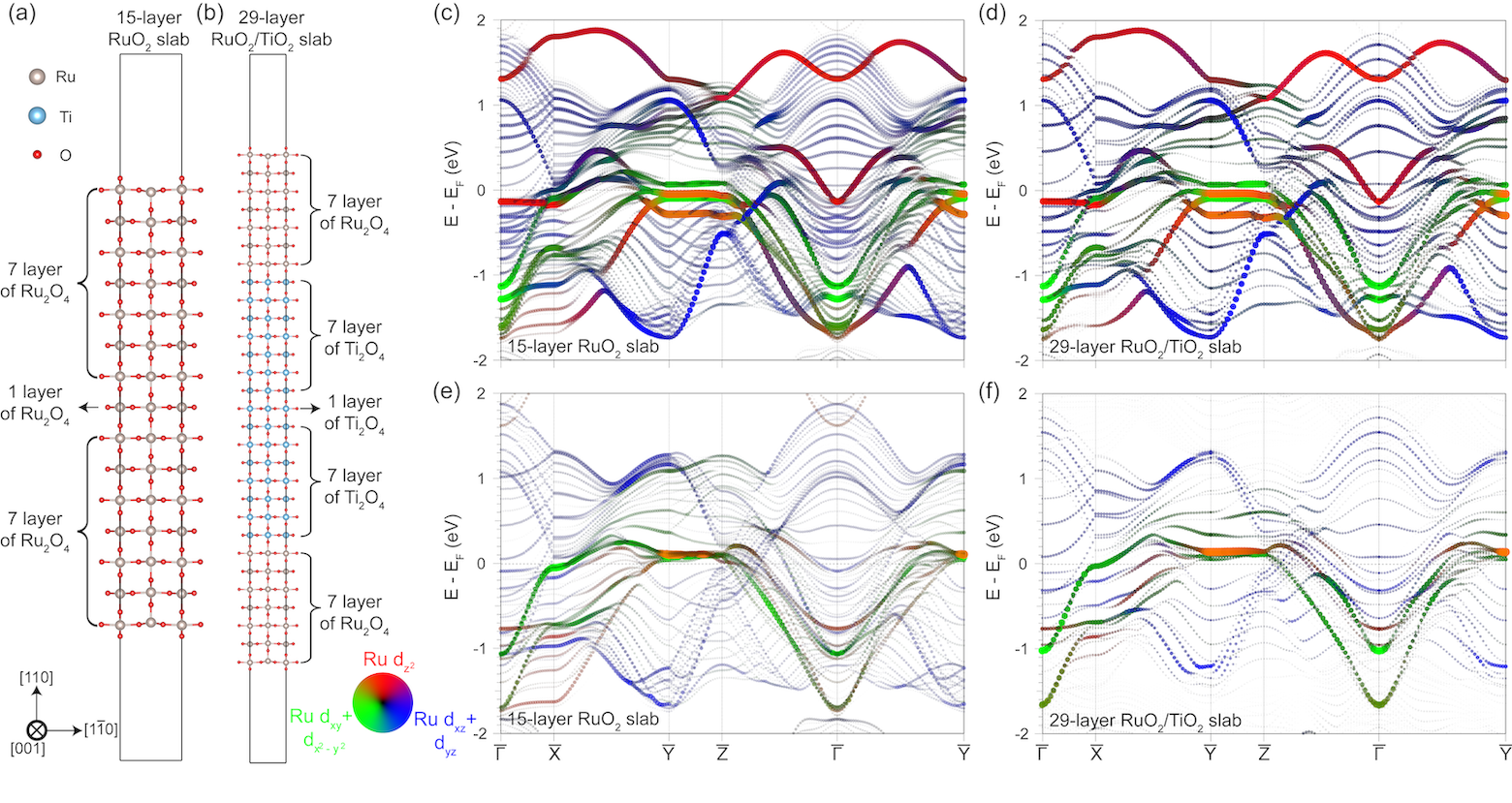}
    \caption{\textbf{Non-spin-polarized density functional theory slab calculations for the inversion symmetric strained RuO$_2$ and RuO$_2$/TiO$_2$ heterostructure.} (\textbf{A} and \textbf{B}) Side views along [001] of the 15-layer strained RuO$_2$ and the 29-layer RuO$_2$/TiO$_2$ slab structures, respectively. The inversion center is contained in the middle layer of RuO$_2$ or TiO$_2$. (\textbf{C}) Electronic band structure of the 15-layer RuO$_2$ under the TiO$_2$ substrate strain where the meaning of the size, transparency, and color-coding of the markers is the same as that of the main text Fig.~\ref{fig:fig3} (D and F). The Brillouin zone notations follow from Fig.~\ref{fig:fig1}C. (\textbf{D}) Same as (C) but for the 29-layer RuO$_2$/TiO$_2$ with the same projection to the surface orbitals. (\textbf{E} and \textbf{F}) Same as (C and D), but the size and transparency of the markers are projected to the summed Ru $d$ and O $p$ orbitals within the 7-th layer counting from the outmost surface layer as the first layer. The red, green, and blue colors also represent the projection to the combinations of the cubic harmonics of the Ru $d$ orbitals within the 7-th layer. The same \ef~shift with Fig.~\ref{fig:fig3} is applied here.}
    \label{fig:fig_nosp_slab_1}
\end{figure}

\begin{figure}
    \centering
    \includegraphics[width=1.0\linewidth]{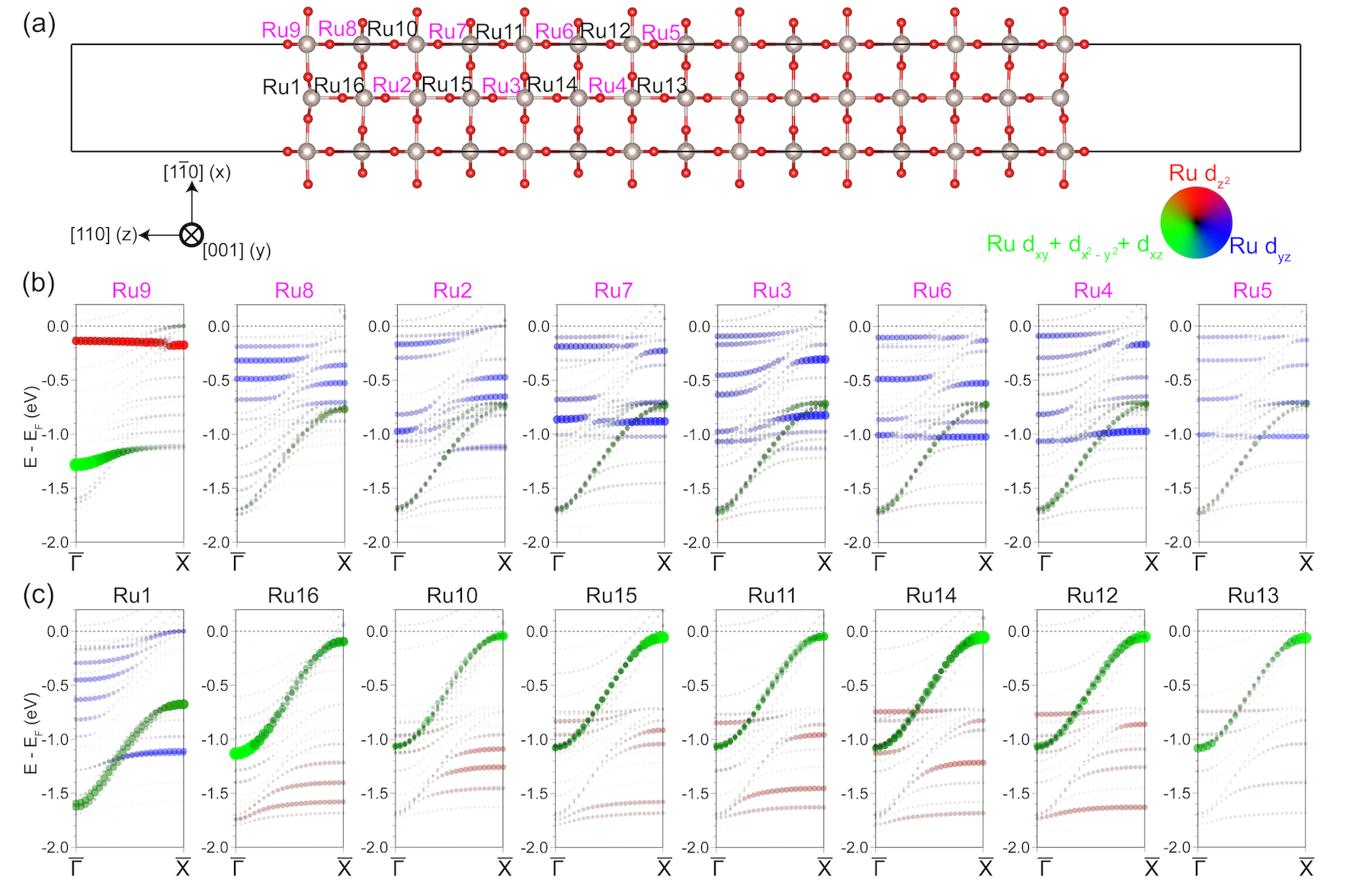}
    \caption{\textbf{Site and orbital-resolved electronic band dispersions of the 15-layer strained RuO$_2$ extracted from non-spin-polarized slab density functional theory calculations.} (\textbf{A}) View of the RuO$_2$ strained slab structure along [001] with numbering on the Ru atoms. The magenta fonts denote Ru atoms that are main contributors to the $\Bar{\Gamma}-\Bar{M}$ narrow bands near the Fermi level. (\textbf{B}) Orbital-resolved band structure along $\Bar{\Gamma}-\Bar{X}$ (half of $\Bar{\Gamma}-\Bar{M}$) projected to the $d$-orbitals of the specified Ru atoms where the red, green, and blue encode the $d_{z^2}$, $d_{xy}+d_{x^2-y^2}+d_{xz}$, and $d_{yz}$ weights of the single Ru atom, respectively. Both the size and transparency of the markers are proportional to the total projection weights onto the specific Ru atoms. (\textbf{C}) Same as (B) but for Ru atoms that do not strongly contribute to $\alpha$-NBs. The same \ef~shift with Fig.~\ref{fig:fig3} is inherited here.}
    \label{fig:fig_nosp_slab_2}
\end{figure}

Further, in order to understand the narrow bands (NBs) along $\Bar{\Gamma}-\Bar{M}$ (termed as $\alpha$-NBs in the main text) while being cautious about the error in estimating Fermi energy and magnetic ground states in slab DFT calculations,  non-spin-polarized slab calculations in Fig.~\ref{fig:fig_nosp_slab_1} and Fig.~\ref{fig:fig_nosp_slab_2} are conducted to approximate the vacuum/RuO$_2$ and RuO$_2$/TiO$_2$ interfaces in experiments. Through our surface annealing procedures described in Materials and Methods of the main text, we are able to obtain a stoichiometric surface termination for the 2.7~nm RuO$_2$ epitaxial layers in ultrahigh vacuum. Therefore, two types of slab structures with stoichiometric terminations are constructed and compared as shown in Fig.~\ref{fig:fig_nosp_slab_1} (A and B), a 15-layer RuO$_2$ slab and a 29-layer RuO$_2$/TiO$_2$ slab, both of which possess inversion symmetry, with the layers pointed out by the horizontal arrows containing the inversion centers. Each stoichiometric layer contains two Ru (or Ti) atoms and four oxygen atoms. Under the parameters specified by the Material and Methods section, both the 15-layer RuO$_2$ and the 29-layer RuO$_2$/TiO$_2$ slab structures yield a calculated work function near 5.31~eV. Then the extracted electronic band structure from the two relaxed slabs along high symmetry directions are shown in Fig.~\ref{fig:fig_nosp_slab_1} (C and D) after projected to orbitals located within the two surface layers. A side-by-side comparison of the two sets of results indicates that the TiO$_2$ substrate influences negligibly the electronic band structure near the surface region under a model of 7 stoichiometric layers of Ru$_2$O$_4$ which has a thickness of roughly 2~nm, close to our ultrathin film measured experimentally. Therefore, the results from the 15-layer strained RuO$_2$ slab are chosen to compare with the surface-sensitive ARPES results in the main text. Next we project the band structure to the 7-th layer in Fig.~\ref{fig:fig_nosp_slab_1} (E and F) to gain insights on effects arising from the RuO$_2$/TiO$_2$ interface under an globally inversion-symmetric structure. The 7-th layer corresponds to the RuO$_2$ layer adjacent to the layer containing the inversion center in the 15-layer RuO$_2$ structure and the RuO$_2$ layer interfaced to TiO$_2$ in the 29-layer RuO$_2$/TiO$_2$ structure. Overall, the in-plane Ru $d$-orbitals (green) are less affected by the TiO$_2$ substrate, while the out-of-plane Ru $d$-orbitals (red and blue) show more prominent changes due to the charge transfer to TiO$_2$. For instance, the $d_{z^2}$ bands dispersing along $\Bar{\Gamma}-\Bar{X}-\Bar{Y}$ roughly from -1.7 eV to -0.2 eV in Fig.~\ref{fig:fig_nosp_slab_1}E show vanishing projected weights in Fig.~\ref{fig:fig_nosp_slab_1}F and the $d_{xz}+d_{yz}$ bands along $\Bar{X}-\Bar{Y}$ near [-1.7, -1.5]~eV relative to the Fermi energy also have altered dispersions. It is worthwhile to point out that the narrow band features near \ef~along $\Bar{\Gamma}-\Bar{X}$ (half of $\Bar{\Gamma}-\Bar{M}$) and $\Bar{Y}-\Bar{Z}$ (half of $\Bar{Z}-\Bar{A}$) do not present a strong contrast between the 15-layer RuO$_2$ and the 29-layer RuO$_2$/TiO$_2$ results even for the projection onto the 7-th layer.

Next in Fig.~\ref{fig:fig_nosp_slab_2}, we study the site and orbital origins of the $\Bar{\Gamma}-\Bar{M}$ narrow bands ($\alpha$-NBs) experimentally observed close to the Fermi level. As shown in Fig.~\ref{fig:fig_nosp_slab_2}A, the strained RuO$_2$ slab is displayed from the view of [001], denoted as the $y$-axis, in accordance with the $k_y$ direction adopted in ARPES measurements. The 16 symmetry nonequivalent Ru atoms are denoted using two colors: magenta and black. The magenta ones are the atoms contributing strongly to $\alpha$-NBs, while the atoms denoted in black provide strong spectral weights for the hole-like dispersive bands topped near the Fermi level at $\Bar{X}$, as can be seen from Fig.~\ref{fig:fig_nosp_slab_2}C. In order to emphasize the distinct orbital characters of $d_{z^2}$ and $d_{yz}$ for the $\alpha$-NBs, the red-green-blue color coding has been divided differently from main text and Fig.~\ref{fig:fig_nosp_slab_1} so that the three Ru $d_{xy}$, $d_{x^2-y^2}$, and $d_{xz}$ are grouped into green. Therefore, a clear trend can be observed in Fig.~\ref{fig:fig_nosp_slab_2}B: the $\alpha$-NBs about 120~meV below the Fermi level consisting of a strong $d_{z^2}$ orbital character of Ru9 at the surface and persisting $d_{yz}$ characters from other Ru atoms marked in magenta extending to the inner layers of the slab.

\begin{figure}
    \centering
    \includegraphics[width=0.5\linewidth]{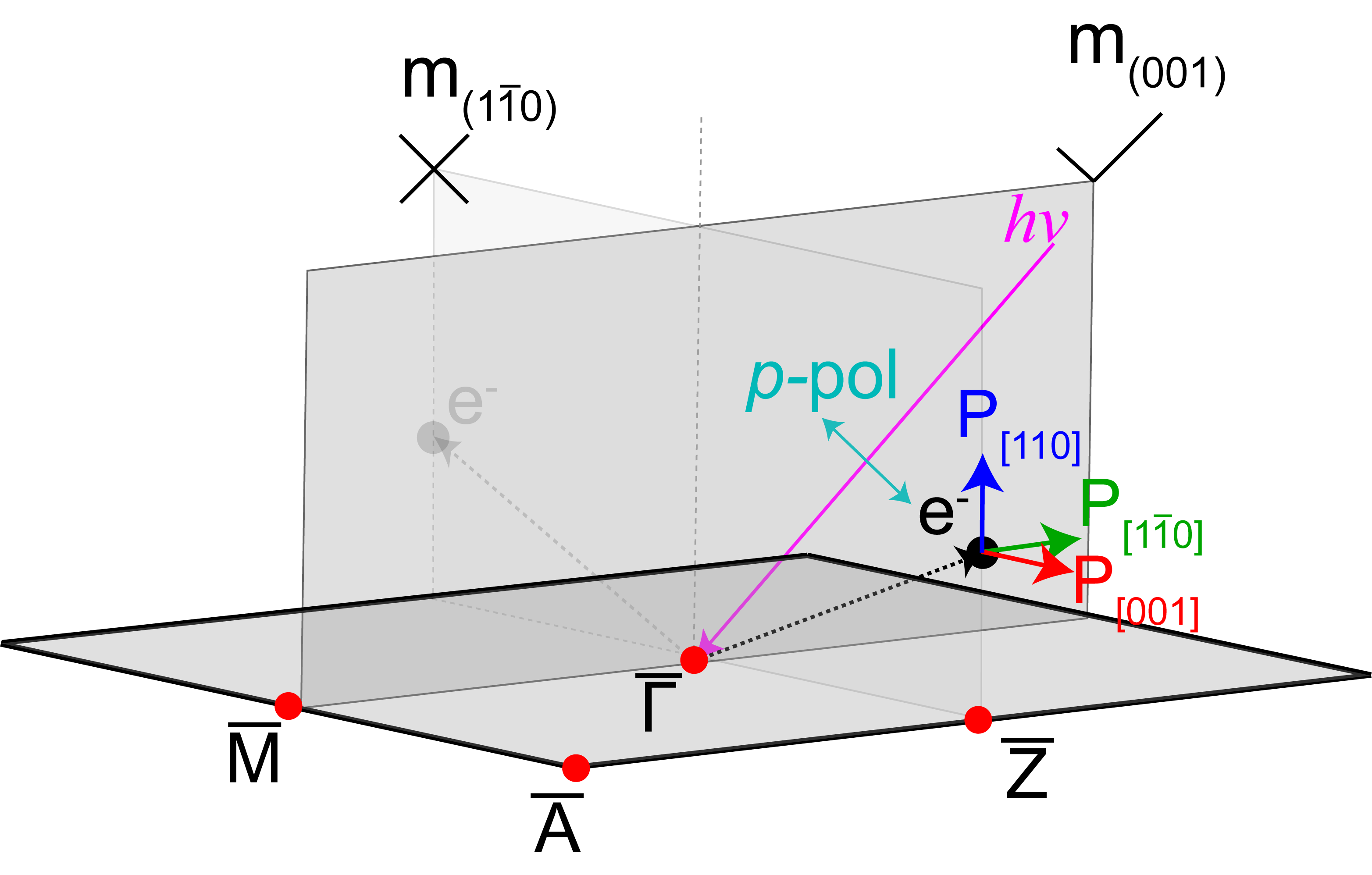}
    \caption{\textbf{Schematic illustration of the photoemission Geometry B.} Similar to Fig.~\ref{fig:fig1}D but with the beam incidence rotated towards within the (001)-mirror plane, therefore preserving the (001)-mirror but breaking the (1$\bar{1}$0)-mirror of the total photoemission system. The three photoelectron spin components are indicated in red, green, and blue arrows.}
    \label{fig:fig_geometry_B}
\end{figure}

\begin{figure}
    \centering
    \includegraphics[width=1.0\linewidth]{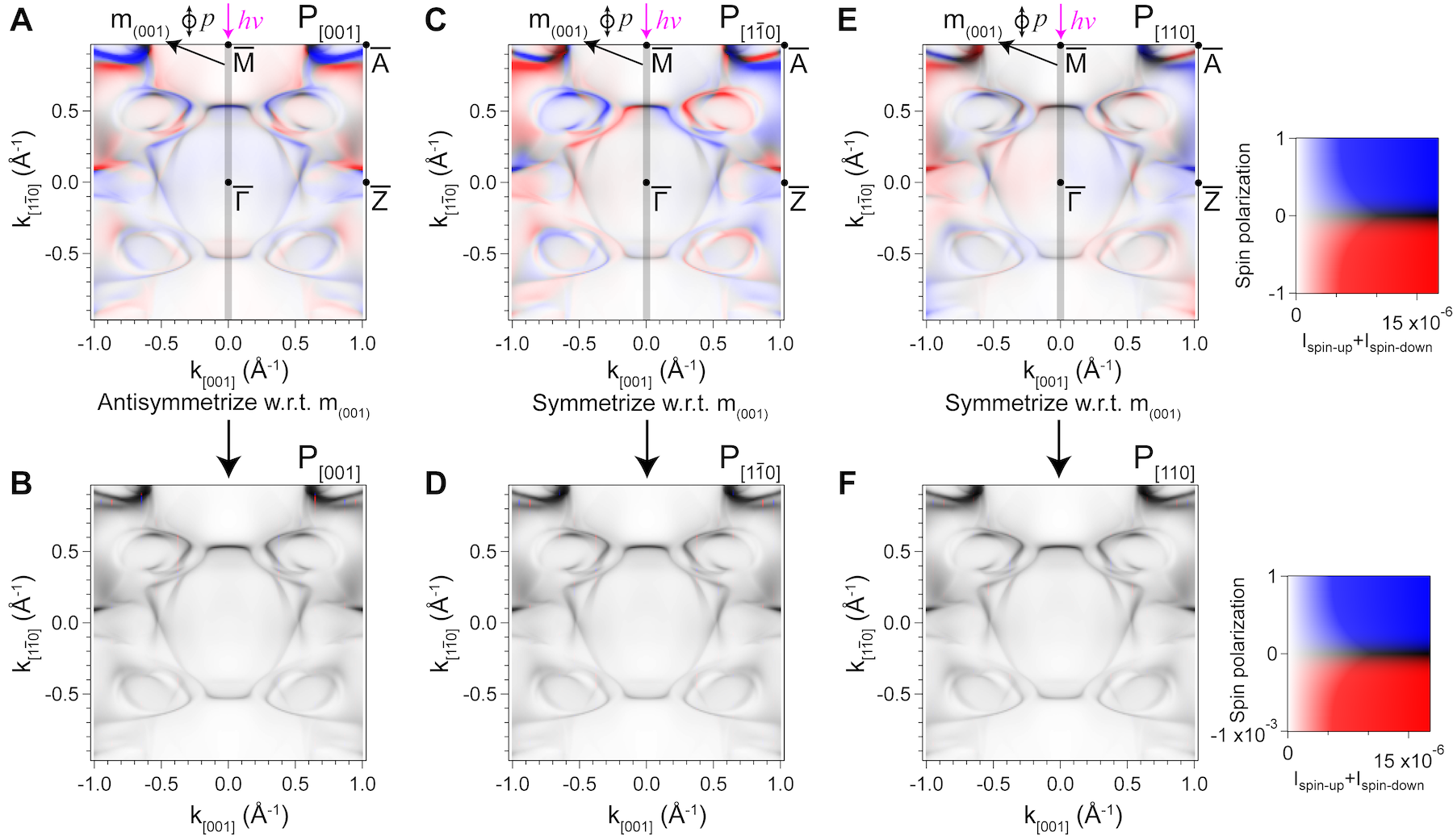}
    \caption{\textbf{Photoelectron spin polarization from fully relativistic first-principles calculations using the one-step model of photoemission on the (110) cleaving surface of nonmagnetic RuO$_2$ under Geometry B and 62~eV $p$-polarized photons.} (\textbf{A}) Calculated photoelectron spin polarization on the Fermi surface projected to the [001] crystalline axis ($P_{[001]}$). Vertical gray bar indicates the preserved (001)-mirror. Light incidence with $p$-polarized photons is indicated on the top, fully within the (001)-mirror. A two-dimensional color scale shown on the right of the top row is employed to faithfully visualize the calculated spin texture. (\textbf{B}) Anti-symmetrizing the results in panel (A) with respect to (w.r.t.) the preserved (001)-mirror. Notice that the corresponding color scale of the spin polarization is zoomed in to $1\times10^{-3}$, as indicated on the right of the bottom row. (\textbf{C} and \textbf{D}) Same as (A and B) but for the photoelectron spin polarization along the [1$\bar{1}$0] direction. (D) symmetrizes the data in (C) w.r.t. the (001)-mirror. (\textbf{E} and \textbf{F}) Same as (A and B) but for the photoelectron spin polarization along the [110] axis. (F) symmetrizes the data in (E) w.r.t. the (001)-mirror.}
    \label{fig:fig_one_step}
\end{figure}

In the last part of first-principles based calculations, we use the fully relativistic one-step model of ARPES to examine the final state selection rules on the photoelectron spin polarization of an assumed time-reversal-symmetry-preserved RuO$_2$. As a proof of principle, we perform the calculations in Geometry B, as sketched in Fig.~\ref{fig:fig_geometry_B}, and numerically confirm the results in main text Table.~\ref{tab:sarpes}. As shown in Fig.~\ref{fig:fig_one_step} (A, C, E), the photoelectron spin polarization defined as $(I_{\rm spin-up}-I_{\rm spin-down})/(I_{\rm spin-up}+I_{\rm spin-down})$ is projected along the [001], [1$\bar{1}$0], and [110] axes, respectively. We find that the spin-polarizations parallel to the (001)-mirror, $P_{[1\bar{1}0]}$ and $P_{[110]}$, display a (001)-mirror-odd texture, while the polarization perpendicular to the mirror, $P_{[001]}$, shows a (001)-mirror-even spin texture. To further validate such conclusions, we anti-symmetrize $P_{[001]}$ and symmetrize $P_{[1\bar{1}0]}$ and $P_{[110]}$ with respect to the (001)-mirror in Fig.~\ref{fig:fig_one_step} (B, D, F). The results, as expected, demonstrate zero values subject to numerical errors. Notice that the two-dimensional color scale for Fig.~\ref{fig:fig_one_step} (B, D, F) is compressed down to $\pm 1\times10^{-3}$.

\subsubsection*{Additional angle-resolved photoemission spectroscopy measurements in spin-integrated and spin-resolved modes}
\begin{figure}
    \centering
    \includegraphics[width=1.0\linewidth]{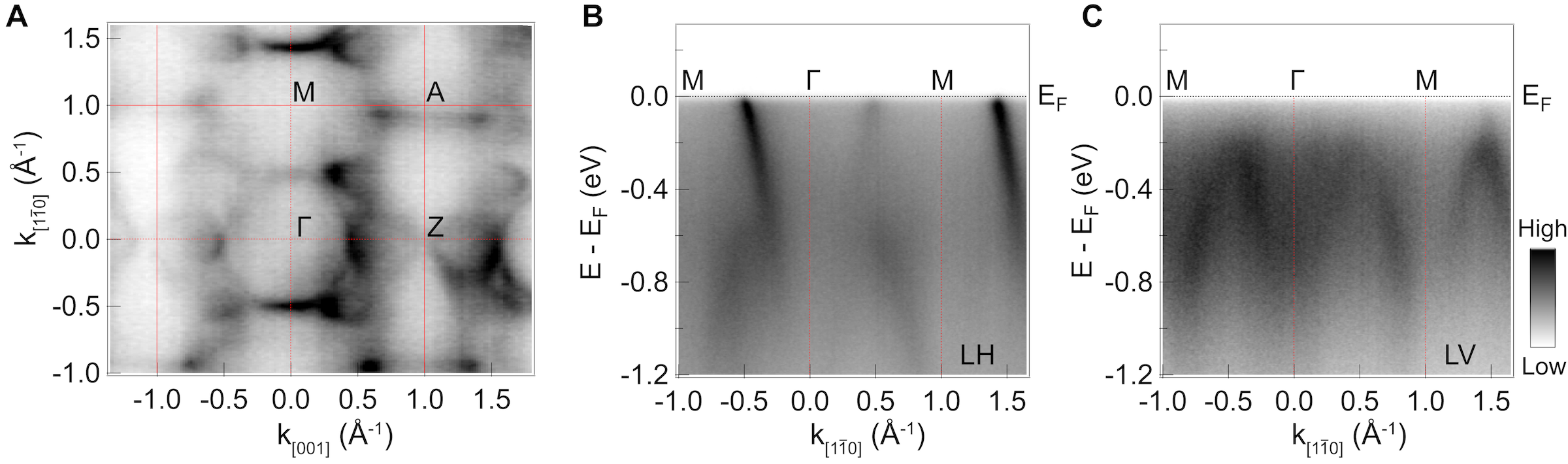}
    \caption{\textbf{Measured electronic band structure of the 14~nm strain-relaxed RuO$_2$ film.} (\textbf{A}) Fermi surface of the (110)-plane. \textbf{(B)} $\Gamma-M$ band dispersions measured with linear horizontal (LH) polarization. (\textbf{C}) $\Gamma-M$ band dispersions measured with linear vertical (LV) polarization. Photons of 130~eV was determined and adopted to measure near the $k_z=0$ plane in the thicker film.}
    \label{fig:fig_14nm_ARPES}
\end{figure}

The angle-resolved photoemission spectroscopy (ARPES) measurements on a partially strain-relaxed 14~nm (110) RuO$_2$/TiO$_2$ film have been conducted to visualize its band structure, as shown in Fig.~\ref{fig:fig_14nm_ARPES}. The 14~nm RuO$_2$ film underwent surface annealing procedure at an elevated temperature of around 560~$^\circ$C after exposure to atmosphere and before ARPES measurements. The measured Fermi surface is displayed in Fig.~\ref{fig:fig_14nm_ARPES}A. Focusing on the $\Gamma-M$ band dispersions using both linear horizontal and linear vertical polarization of light in Supplementary Fig.~\ref{fig:fig_14nm_ARPES} (B and C), we do not observe any features associated with the $\alpha$-NBs, although other dispersive bands exhibit certain resemblance to the data in main text Fig.~\ref{fig:fig2} (C and D). This indicates that the $\alpha$-NBs could be sensitive to post-annealing temperatures, disappearing due to surface oxygen vacancies at this elevated annealing temperature.

\begin{figure}
    \centering
    \includegraphics[width=1.0\linewidth]{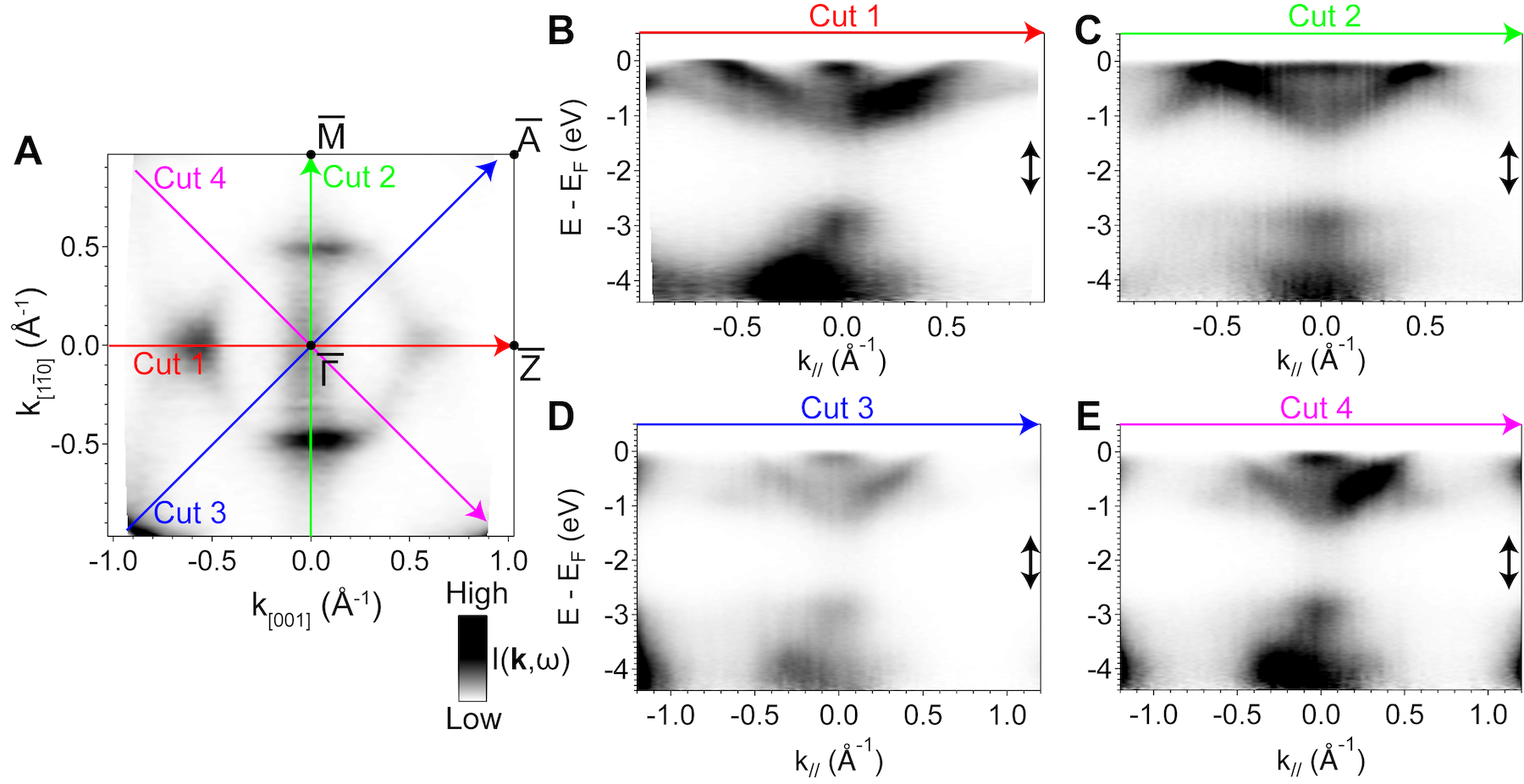}
    \caption{\textbf{Valence bands of the epitaxially strained RuO$_2$ across a larger binding energy range.} (\textbf{A}) Measured Fermi surface similar to main text Fig.~\ref{fig:fig4}A with 62~eV $p$-polarized photons, but with all scans using a coarser energy step. Notice that the color scale for all plots here is enhanced by a same factor of 2 to emphasize regions with weak spectral intensity, as indicated by the color bar beneath panel (A). (\textbf{B}, \textbf{C}, \textbf{D}, \textbf{E}) Extracted electronic band dispersions along Cut 1, 2, 3, 4, respectively, as indicated by the colored arrows on panel (A). The vertical black double arrows in (B to E) indicate a region with suppressed density of states.}
    \label{fig:fig_deeperE_ARPES}
\end{figure}

Regarding the fully epitaxially strained 2~nm RuO$_2$ films, additional ARPES and spin-resolved ARPES measurements have been carried out. In the spin-integrated mode, a $E-k_x-k_y$ mapping has been scanned down to deeper binding energies in Fig.~\ref{fig:fig_deeperE_ARPES} compared to those measured in the main text. Band dispersions along four representative high symmetry directions are extracted to show in Fig.~\ref{fig:fig_deeperE_ARPES} (B to E), where a strong suppression of density of states between 1.7 and 2.4~eV binding energies (indicated by the vertical black double arrows) is observed across the whole data cube. This is an important indication from the valence bands that the measured RuO$_2$ (110) surface is close to stoichiometry, neither towards oxygen rich, nor oxygen deficient~\cite{Jovic2021}. According to a detailed study~\cite{Jovic2021} on the valence band behavior of single-crystal RuO$_2$ (110) surface under different surface adsorption conditions combining ARPES and DFT, if the surface were oxygen-rich, along the equivalent in-plane momentum path of $\bar{\Gamma}-\bar{M}$ (see Fig.~\ref{fig:fig_deeperE_ARPES}C), a set of valence bands topping around $E-E_{\rm F}=-2.5$~eV would rise up to $E-E_{\rm F}=-2$~eV. On the other hand, if the surface were oxygen-deficient, the $\alpha$-NBs would disappear. However, neither case matches with our observed band dispersions. The persistence of the slightly shifted $\alpha$-NBs due to epitaxial strain and the strongly suppressed spectral weight between $E-E_{\rm F}=-2.4$ and -1.7~eV is fully consistent with the stoichiometric surface~\cite{Jovic2021}.

\begin{figure}
    \centering
    \includegraphics[width=1.0\linewidth]{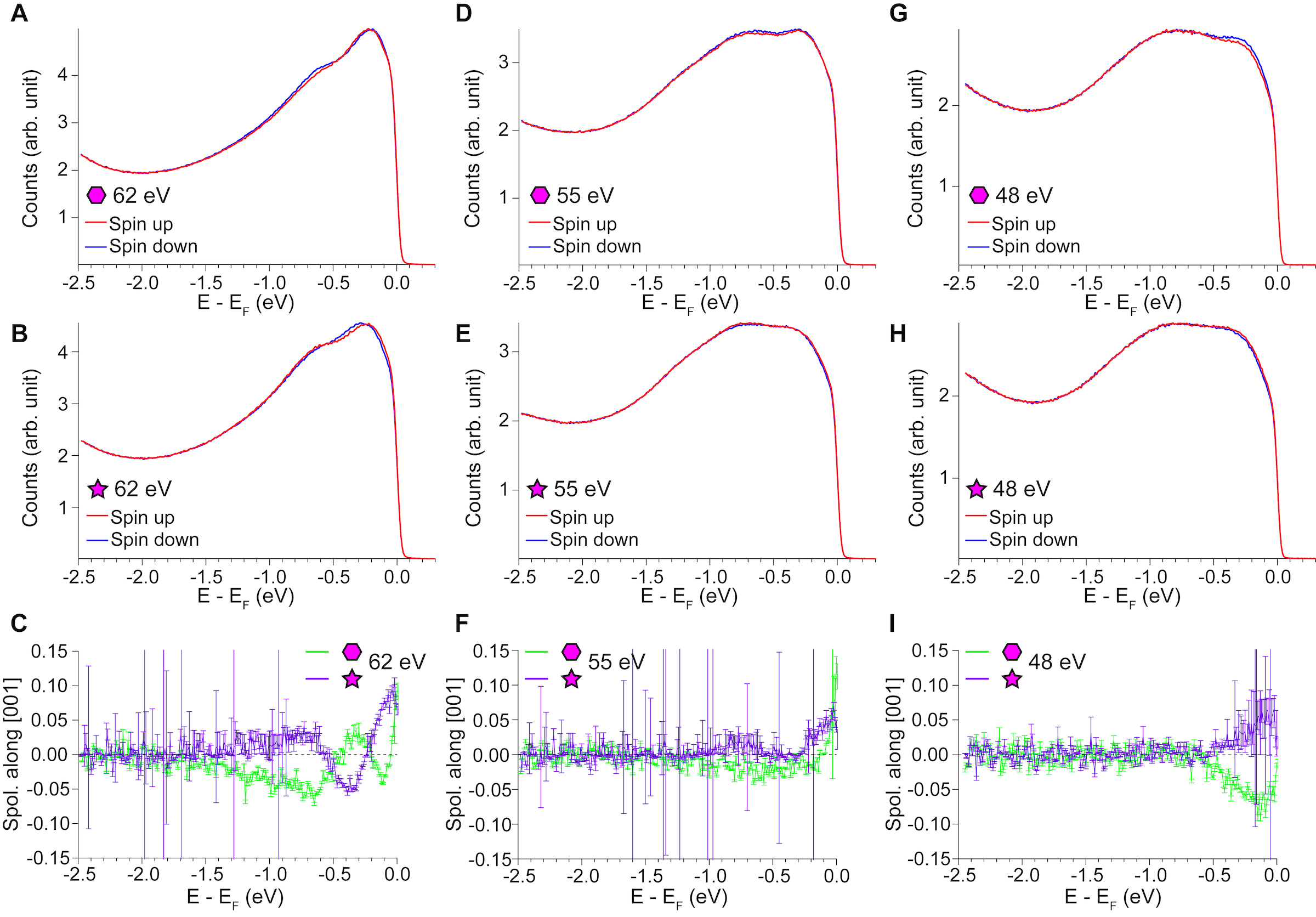}
    \caption{\textbf{Photon energy dependence for the finite-momenta spin-resolved feature in main text Fig.~\ref{fig:fig4} (A to F).} (\textbf{A} and \textbf{B}) Spin-resolved energy distribution curves (EDCs) reproduced from Fig.~\ref{fig:fig4} (C and D). (\textbf{C}) Photoelectron spin polarization along [001] reproduced from Fig.~\ref{fig:fig4} (E and F) but now overlaid on a single plot. The photoemission angles chosen for panel (A to C) are from $+7.5^\circ$ and $-7.5^\circ$. (\textbf{D}, \textbf{E}, \textbf{F}) Counterparts of (A to C) at 55~eV photon energy. Therefore, the photoelectron collection angle is moved to $\pm8.0^\circ$ to keep the same in-plane momenta on the Fermi surface. (\textbf{G}, \textbf{H}, \textbf{I}) Counterparts of (A to C) at 48~eV photon energy, with the angle of photoelectron collection increased to $\pm8.8^\circ$.}
    \label{fig:fig_geometry_A_GM_S001_hv_dep}
\end{figure}

\begin{figure}
    \centering
    \includegraphics[width=1.0\linewidth]{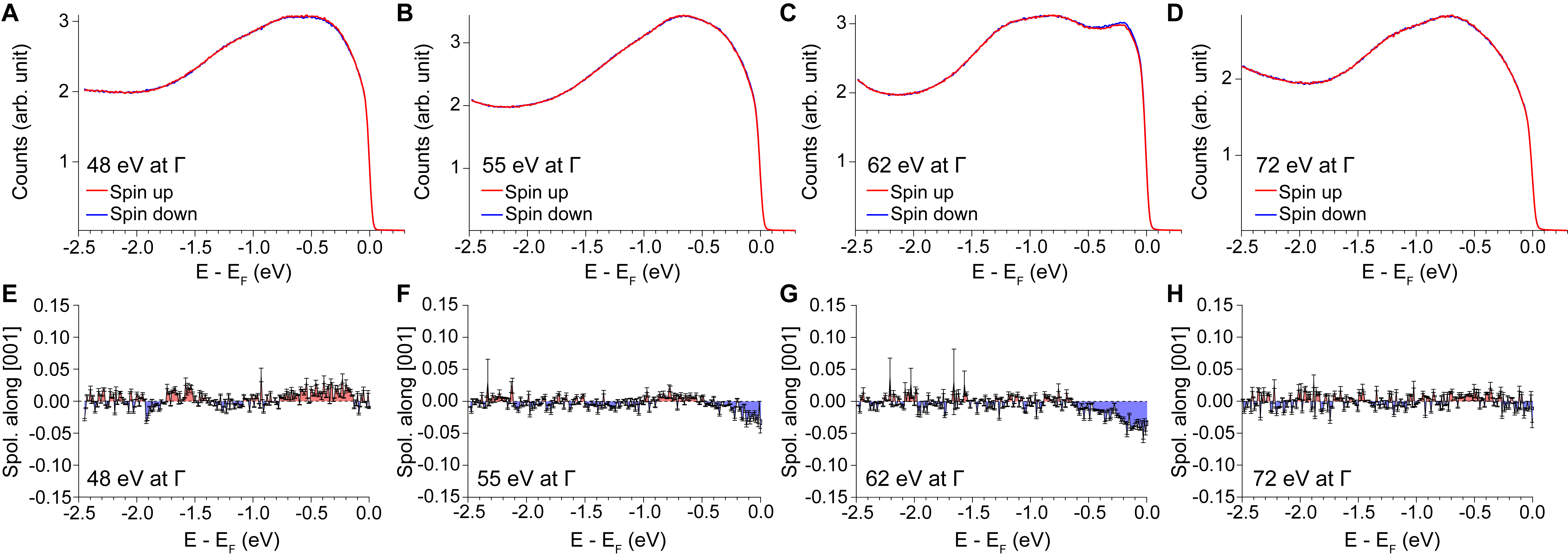}
    \caption{\textbf{Photon energy dependence for the normal emission spin-resolved feature discussed in main text Fig.~\ref{fig:fig4} (G and H).} (\textbf{A}, \textbf{B}, \textbf{C}, \textbf{D}) Spin-resolved energy distribution curves measured at normal emission selecting photoelectrons polarized along the [001] direction at photon energies  of 48, 55, 62, and 72~eV, respectively. (\textbf{E}, \textbf{F}, \textbf{G}, \textbf{H}) Calculated spin polarization $P_{[001]}$ at normal emission based on (A to D).}
    \label{fig:fig_geometry_A_G_S001_hv_dep}
\end{figure}

In the spin-resolved mode, we further investigate the photon energy dependence of the (1$\bar{1}$0)-mirror-even behavior of $P_{[001]}$ discussed in main text Fig.~\ref{fig:fig4}. It is important to note that if the underlying system is time-reversal-symmetry (TRS)-preserved (nonmagnetic or paramagnetic), the final state selection rules imposed by main text Table.~\ref{tab:sarpes} apply to all photon energies, regardless of the microscopic details and the multiple scattering of photoelectrons. After reproducing the results of Fig.~\ref{fig:fig4} (C to F) in Fig.~\ref{fig:fig_geometry_A_GM_S001_hv_dep} (A to C), we also show the $P_{[001]}$ measured at the same in-plane momenta but with photon energies of 55 and 48~eV (Fig.~\ref{fig:fig_geometry_A_GM_S001_hv_dep} (D to F) and (G to I)). The quantitative spin polarization alters as $h\nu$ varies, while at 55~eV the (1$\bar{1}$0)-mirror-even component near $E_{\rm F}$ clearly persists. However, at 48~eV the mirror-even component appears dominated by the mirror-odd component. These observations further substantiate the fact that the photoelectron spin polarization does not necessarily reflect the pre-emission spin texture of the Bloch electrons, due to the relativistic effects and multiple scattering effects that must be considered during photoemission. The positive argument is that as long as the final state selection rules are violated and can be observed at certain photon energies, even though not across all photon energies, then there must be symmetry breaking. And in this case, we attribute it to time-reversal-symmetry broken magnetic states. The same argument can be applied to the photon energy dependence  of the normal emission finite $P_{[001]}$ shown in Fig.~\ref{fig:fig_geometry_A_G_S001_hv_dep}.

\begin{figure}
    \centering
    \includegraphics[width=0.9\linewidth]{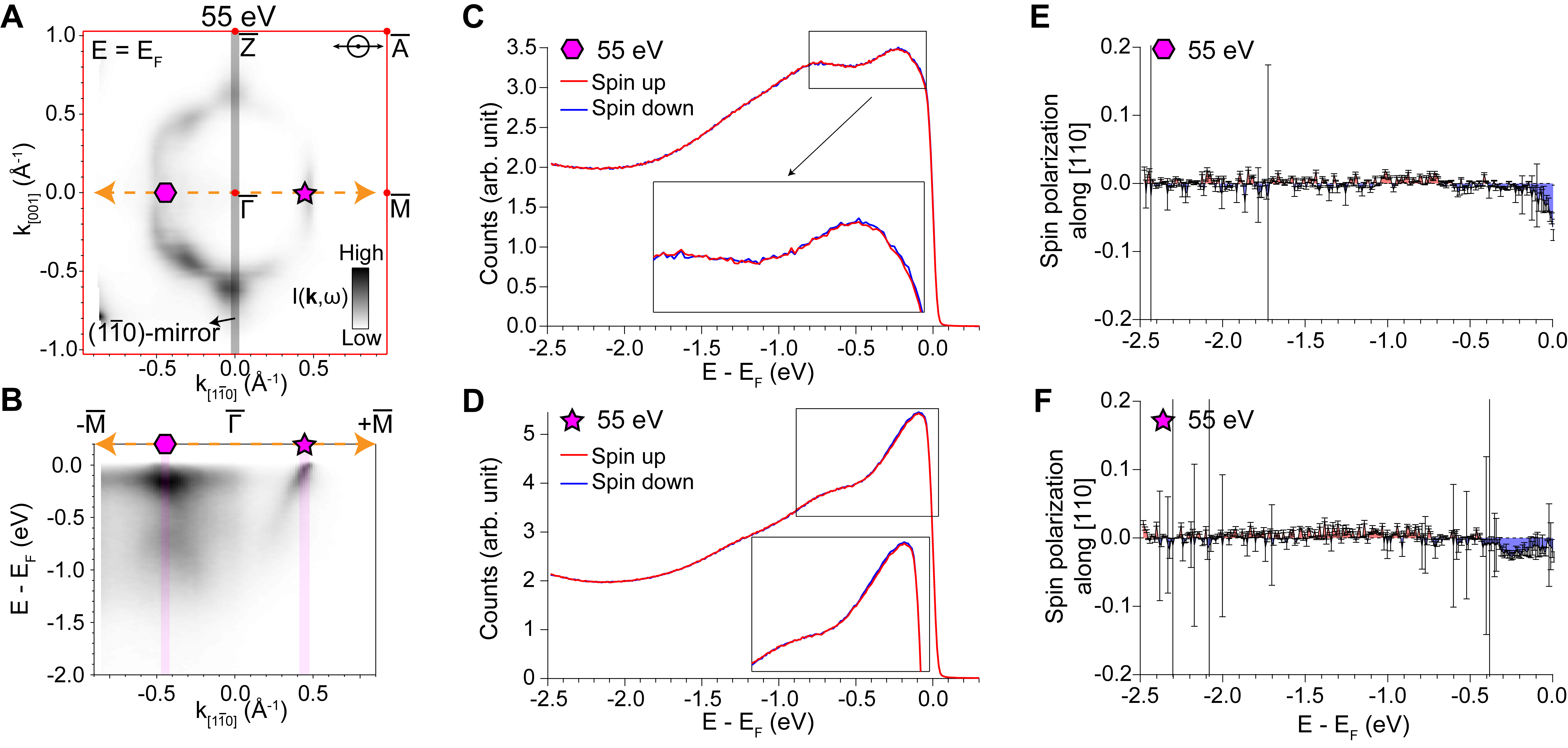}
    \caption{\textbf{Measured angle-resolved photoemission spectroscopy maps and photoelectron spin polarization along the out-of-plane [110] direction on the $\Bar{\Gamma}-\Bar{M}$ path at 55 eV with respect to the broken (1$\mathbf{\bar{1}}$0)-mirror under Geometry B.} (\textbf{A}) Fermi surface probed by 55~eV photons. (\textbf{B}) Band dispersions along the $\bar{\Gamma}-\bar{M}$ direction indicated by the horizontal dashed double-arrow in (A). (\textbf{C} and \textbf{D}) Spin-resolved energy distribution curves integrated across the magenta bars on either sides of the (1$\bar{1}$0)-mirror in (B), selecting photoelectrons with spin polarization only along the out-of-plane [110] direction. (\textbf{E} and \textbf{F}) Converted spin polarization from (C and D), respectively.}
    \label{fig:fig_Geometry_B_GM_55eV_S110}
\end{figure}

Next we present the spin-resolved ARPES data measured at momenta that are not perpendicular to the mirror plane of the total photoemission system. First, in Geometry B, the out-of-plane spin polarization $P_{[110]}$ on specific momenta along $\bar{\Gamma}-\bar{M}$ is measured with 55~eV photons, as marked in Fig.~\ref{fig:fig_Geometry_B_GM_55eV_S110} (A and B) on both the Fermi surface and $\bar{\Gamma}-\bar{M}$ band dispersion plots. The (1$\bar{1}$0)-mirror indicated by the gray bar in Fig.~\ref{fig:fig_Geometry_B_GM_55eV_S110}A, although being a symmetry of the RuO$_2$ film, is no longer a preserved mirror due to the light incidence from right to left on Fig.~\ref{fig:fig_Geometry_B_GM_55eV_S110}A. The measured [110] spin-resolved energy distribution curves (EDCs) and their associated spin polarization are displayed in Fig.~\ref{fig:fig_Geometry_B_GM_55eV_S110} (C to F), showing a negligible value compared to its in-plane spin polarization shown in main text Fig.~\ref{fig:fig4} (E and F). Therefore, it is likely that the initial-state spin polarization near $k_{//}=\pm0.5\textup{~\AA}^{-1}$ on $\bar{\Gamma}-\bar{M}$ is predominantly in-plane. Since the (1$\bar{1}$0)-mirror is not preserved under photoemission, the detected photoelectron spin polarization is strongly affected by the initial-state Bloch electron spin polarization and final-state effects. 

\begin{figure}
    \centering
    \includegraphics[width=1.0\linewidth]{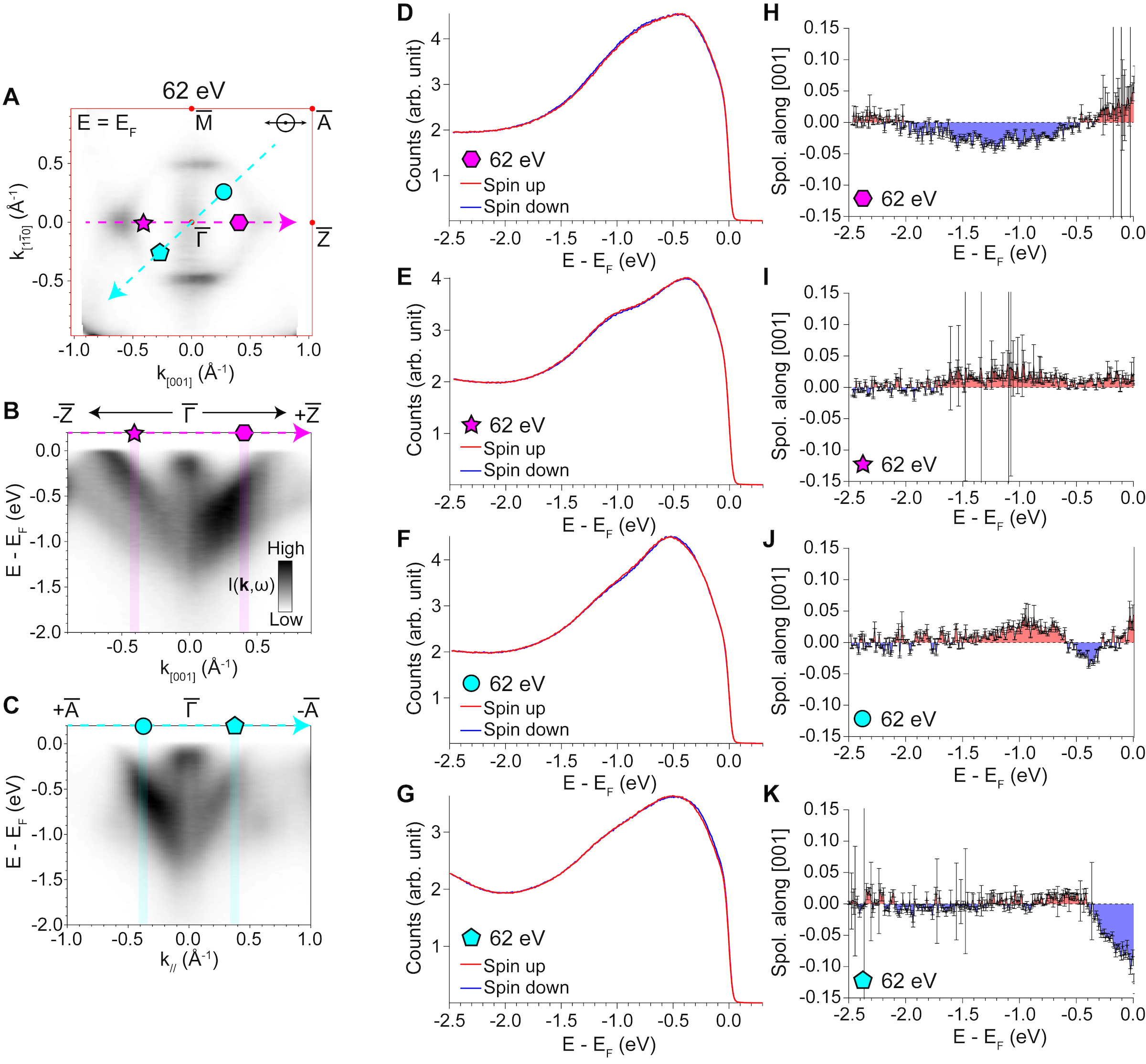}
    \caption{\textbf{Measured band dispersions and in-plane [001] photoelectron spin polarization on the $\Bar{\Gamma}-\Bar{Z}$ and $\Bar{\Gamma}-\Bar{A}$ paths using 62~eV light under Geometry A.} (\textbf{A}) Fermi surface reproduced from Fig.~\ref{fig:fig4}A. Magenta and cyan symbols indicate where the spin-resolved energy distribution curves (EDCs) are taken on $\Bar{\Gamma}-\Bar{Z}$ and $\Bar{\Gamma}-\Bar{A}$ lines. (\textbf{B} and \textbf{C}) Electronic band dispersions extracted along $\Bar{\Gamma}-\Bar{Z}$ and $\Bar{\Gamma}-\Bar{A}$, as indicated by the directional magenta and cyan arrows in panel (A). (\textbf{D}, \textbf{E}, \textbf{F}, \textbf{G}) Spin-resolved EDCs for the [001] photoelectrons spins measured at the four momenta indicated on (A). (\textbf{H}, \textbf{I}, \textbf{J}, \textbf{K}) Converted photoelectron spin polarization $P_{[001]}$ based on the data in (D to G).}
    \label{fig:fig_geometry_A_GZ_diag_62eV_S001}
\end{figure}

\begin{figure}
    \centering
    \includegraphics[width=1.0\linewidth]{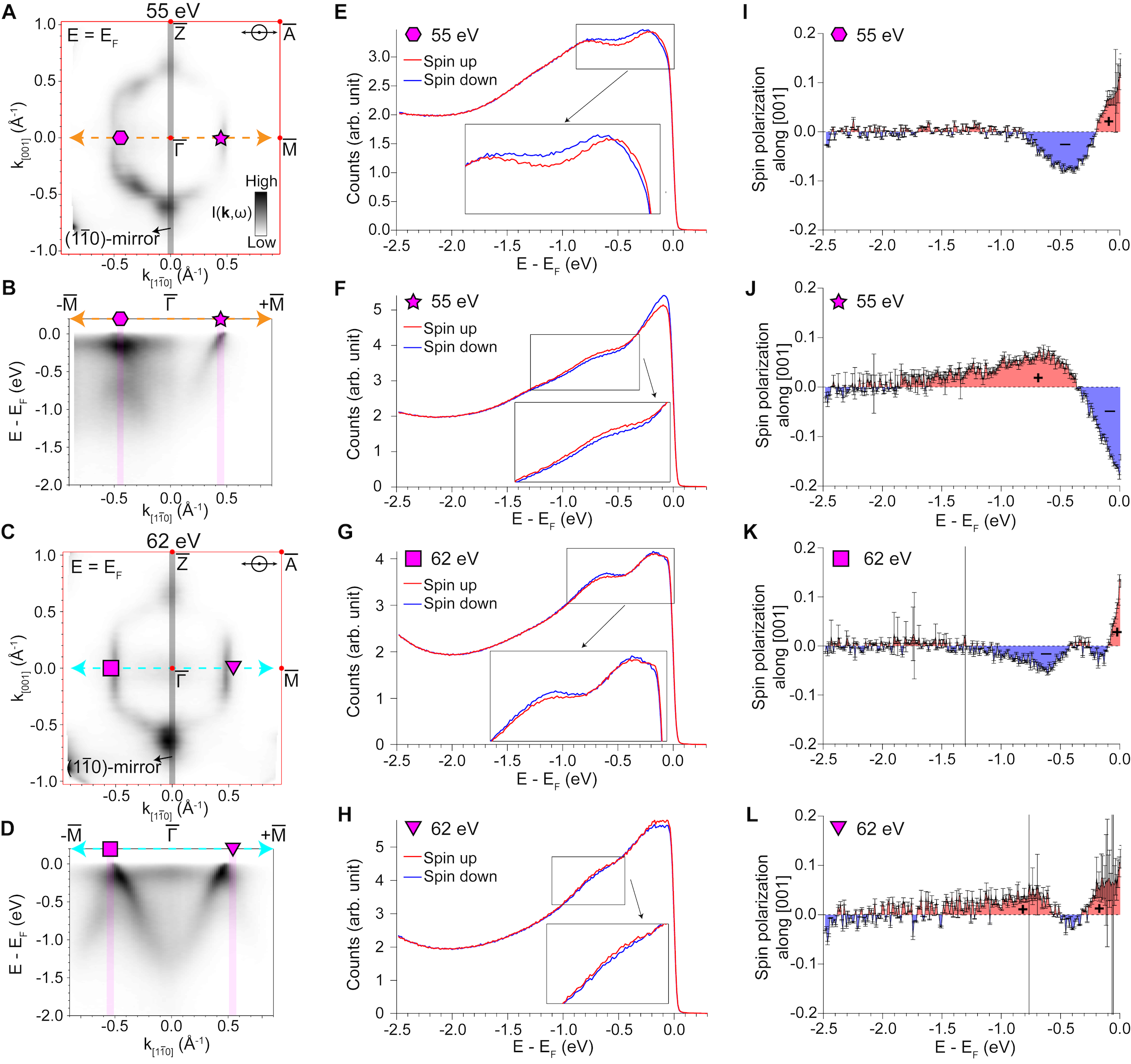}
    \caption{\textbf{Measured photoelectron spin polarization along the in-plane [001] direction on the $\Bar{\Gamma}-\Bar{M}$ path with respect to the broken (1$\mathbf{\bar{1}}$0)-mirror plane under Geometry B.} (\textbf{A} and \textbf{B}) Fermi surface and band dispersions reproduced from Fig.~\ref{fig:fig_Geometry_B_GM_55eV_S110} (A and B). (\textbf{C} and \textbf{D}) Same as (A and B), but taken using the 62~eV photon energy. The magenta symbols in (A to D) indicate where the spin-resolved energy distribution curves (EDCs) are measured. (\textbf{E} and \textbf{F}) Raw spin-resolved EDCs integrated across the magenta bars on either sides of the (1$\bar{1}$0)-mirror in panel (B), selecting photoelectrons with spin polarization only along the in-plane [001] direction. (\textbf{G} and \textbf{H}) Same as (E and F), but for 62~eV. (\textbf{I}, \textbf{J}, \textbf{K}, \textbf{L}) Converted spin polarization based on (E, F, G, H), respectively.}
    \label{fig:fig_old_main_4}
\end{figure}

Further measured spin-resolved ARPES data along $\Bar{\Gamma}-\Bar{Z}$, $\Bar{\Gamma}-\Bar{A}$, and $\Bar{\Gamma}-\Bar{M}$ are displayed in Fig.~\ref{fig:fig_geometry_A_GZ_diag_62eV_S001} (under Geometry A) and Fig.~\ref{fig:fig_old_main_4} (under Geometry B) for in-plane $P_{[001]}$ spin polarization. As shown in Fig.~\ref{fig:fig_geometry_A_GZ_diag_62eV_S001} (A to C) and Fig.~\ref{fig:fig_old_main_4} (A to D) the $P_{[001]}$ spin polarization in general shows a mixture of mirror-odd and mirror-even behavior, as evidenced by the datasets in Fig.~\ref{fig:fig_geometry_A_GZ_diag_62eV_S001} (H to K) and Fig.~\ref{fig:fig_old_main_4} (I to L). Similar considerations to those presented above in the context of Fig. \ref{fig:fig_Geometry_B_GM_55eV_S110} apply here, since the (001)-mirror is not preserved in the Geometry A and the (1$\bar{1}$0)-mirror is not preserved in Geometry B. The specific extrinsic mechanisms that determine the photoelectron spin polarization in these cases are beyond the focus of the current work. One final remark for this paragraph is that the relatively large $P_{[001]}$ polarization showcased in Fig.~\ref{fig:fig_old_main_4} (I to L) at both 55 and 62~eV supports the conclusion of predominant in-plane spin polarization along $\bar{\Gamma}\bar{M}$ near the $k_{//}=\pm0.5\textup{~\AA}^{-1}$ region.

\begin{figure}
    \centering
    \includegraphics[width=1.0\linewidth]{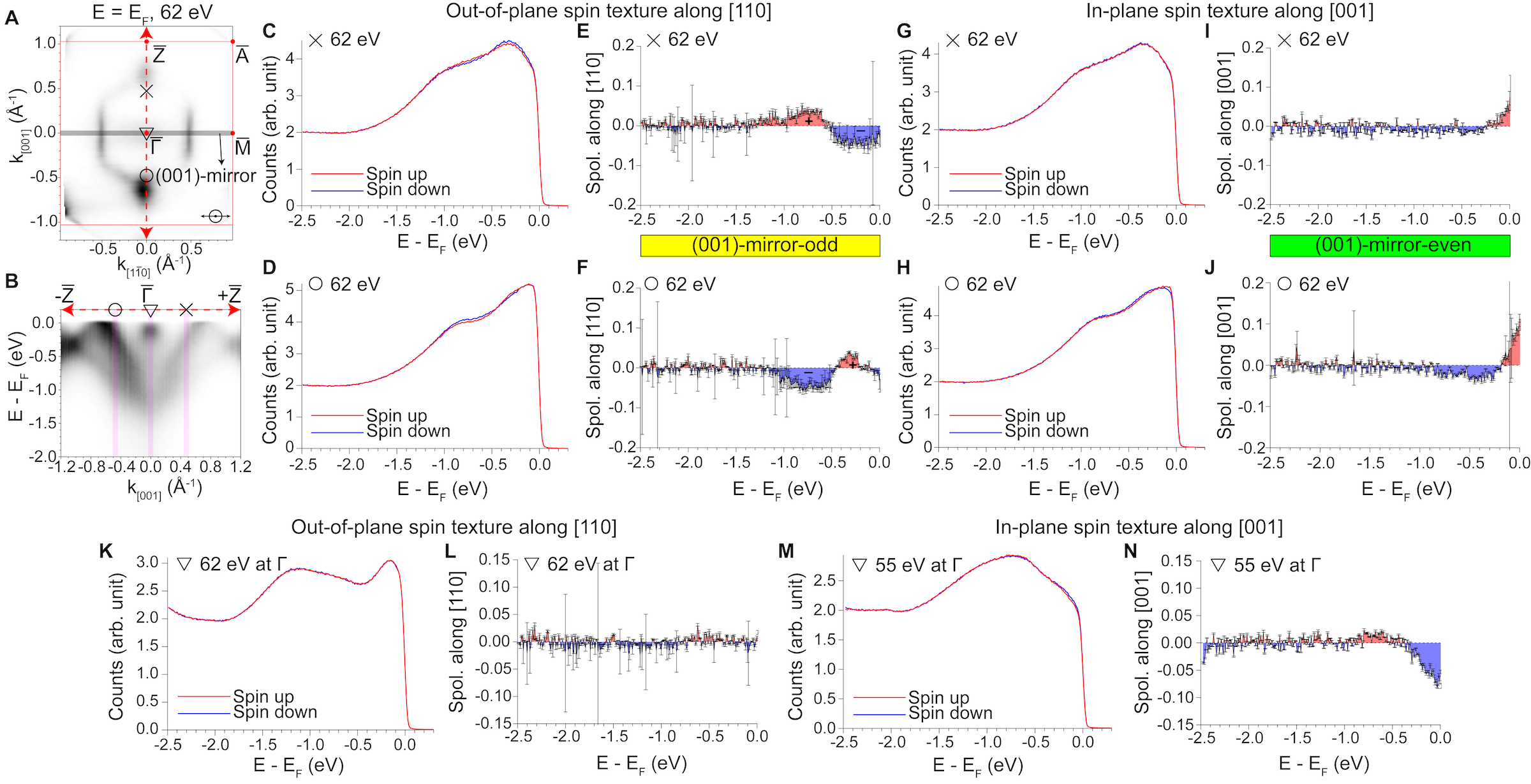}
    \caption{\textbf{Out-of-plane [110] and in-plane [001] photoelectron spin polarization on the $\Bar{\Gamma}-\Bar{Z}$ momentum path with respect to the preserved (001)-mirror plane measured under Geometry B.} (\textbf{A}) Fermi surface reproduced from Fig.~\ref{fig:fig_old_main_4}C but highlighting the (001)-mirror and the momentum positions of the measured spin-resolved energy distribution curves (EDCs) using circular, cross, and triangular symbols. (\textbf{B}) Electronic band dispersions measured along $\bar{\Gamma}-\bar{Z}$ at 62 eV. The width of the vertical magenta bars provides an estimation of the momentum resolution in the spin-resolved measurement mode. (\textbf{C} and \textbf{D}) Spin-resolved EDCs selectively probing only the out-of-plane [110] spin polarization on the upper and lower sides of the (001)-mirror. (\textbf{E} and \textbf{F}) Converted out-of-plane spin polarization from (C and D), respectively. (\textbf{G}, \textbf{H}, \textbf{I}, \textbf{J}) Same as (C, D, E, F), but probing the in-plane spin polarization along [001]. (\textbf{K}) Spin-resolved EDCs selecting the out-of-plane [110] axis measured at normal emission using 62~eV light. (\textbf{L}) Converted [110] photoelectron spin polarization from (K). (\textbf{M} and \textbf{N}) Same as (K and L) but for the in-plane [001] spin quantization axis at normal emission using 55~eV photons.}
    \label{fig:fig_Geometry_B_GZ_G_62eV_55eV_S110_S001}
\end{figure}

In the following, we present more comprehensive measurements to support the conclusions drawn from main text Fig.~\ref{fig:fig5} measured under Geometry B and with the sample magnetized by an out-of-plane magnetic field of 0.4~T. As shown in Fig.~\ref{fig:fig_Geometry_B_GZ_G_62eV_55eV_S110_S001}A, the (001)-mirror is a preserved mirror under Geometry B. Therefore, along the $\bar{\Gamma}-\bar{Z}$ path, final state selection rules from the second row of main text Table.~\ref{tab:sarpes} would apply if the system were TRS-preserved. Experimentally, as shown in Fig.~\ref{fig:fig_Geometry_B_GZ_G_62eV_55eV_S110_S001} (C to F) and (K and L), we observe the persistent (001)-mirror-odd behavior and a zero spin polarization at normal emission for $P_{[110]}$ at 62~eV, in addition to the consistent observations presented in main text Fig.~\ref{fig:fig5} for 55~eV measurements. Further, we measured the in-plane $P_{[001]}$  spin polarization, which is allowed to be non-zero in this geometry by the ARPES selection rules even if the material does not intrinsically break TRS. As shown in Fig.~\ref{fig:fig_Geometry_B_GZ_G_62eV_55eV_S110_S001} (G to J) and (M and N), $P_{[001]}$ shows a (001)-mirror-even behavior and a finite spin polarization at normal emission. Therefore, this observed $P_{[001]}$ spin splitting cannot be used as unambiguous evidence of symmetry breaking, unlike the ones carried out in main text Fig.~\ref{fig:fig4} with the sample magnetized by an in-plane field of 0.2~T. These controlled experiments further support the interpretation that the (1$\bar{1}$0)-mirror-even components of $P_{[001]}$ observed in main text Fig.~\ref{fig:fig4} under Geometry A are associated with the intrinsic TRS breaking occurring within the [001] spin channel.

\begin{figure}
    \centering
    \includegraphics[width=1.0\linewidth]{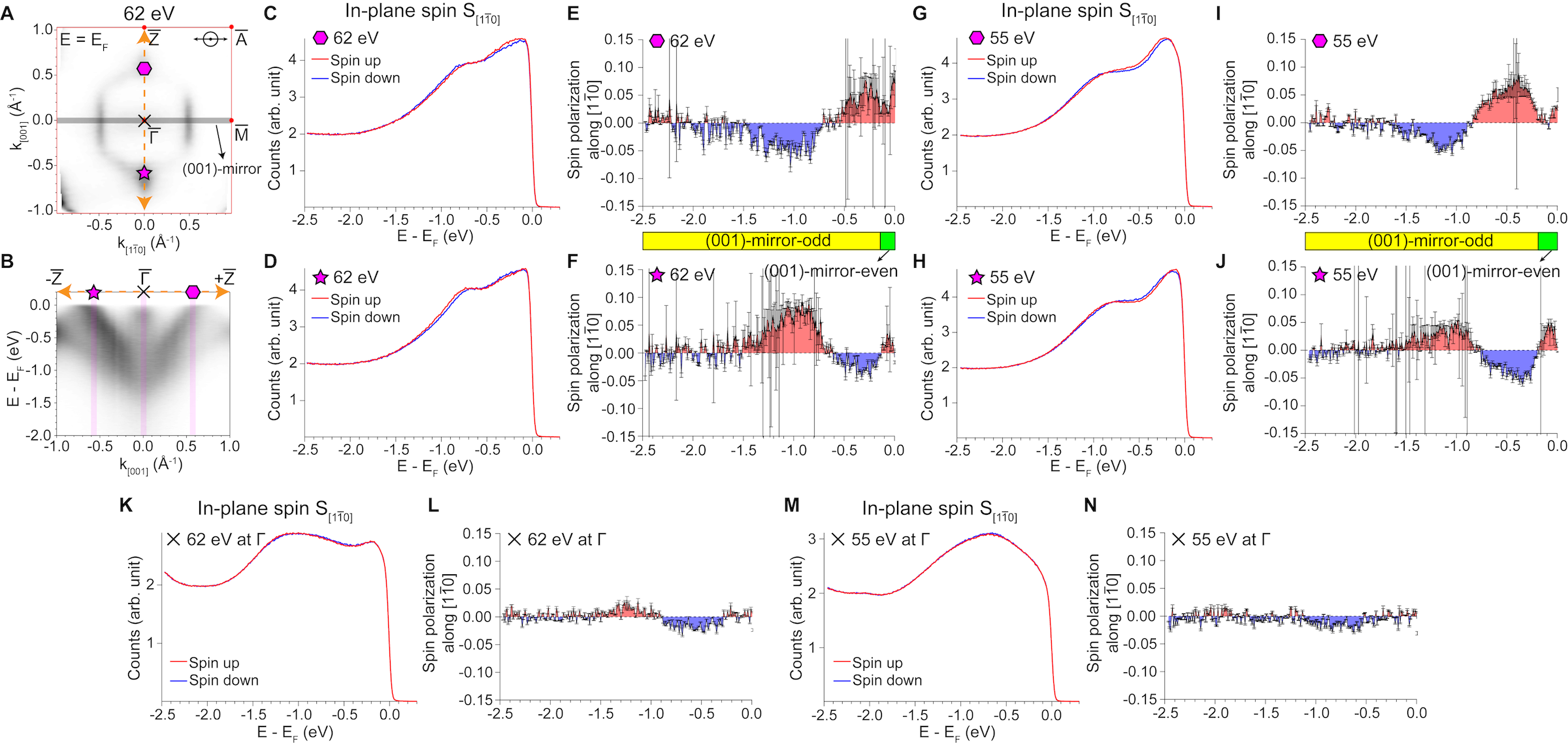}
    \caption{\textbf{Measured photoelectron spin polarization along the in-plane [1$\mathbf{\bar{1}}$0] axis on the $\Bar{\Gamma}-\Bar{Z}$ path with respect to the preserved (001)-mirror under Geometry B.} (\textbf{A}) Measured Fermi surface at 62 eV, which comes from the same sample measured in Fig.~\ref{fig:fig_geometry_A_GZ_diag_62eV_S001}A but rotated azimuthally 90$^\circ$. (\textbf{B}) Extracted $\Bar{\Gamma}-\Bar{Z}$ band structure. (\textbf{C} and \textbf{D}) [1$\bar{1}$0] spin-resolved energy distribution curves (EDCs) measured at the momenta indicated by the pentagon and asterisk symbols respectively, mirror symmetric with respect to the preserved (001)-mirror. (\textbf{E} and \textbf{F}) Calculated $P_{[1\bar{1}0]}$ from (C and D). (\textbf{G} to \textbf{J}) Same as (C to F) but measured with 55~eV photons, while keeping the in-plane momenta near the Fermi level the same. (\textbf{K}) [1$\bar{1}$0] spin-resolved EDCs at normal emission under 62~eV light. (\textbf{L}) $P_{[1\bar{1}0]}$ from data in (K). (\textbf{M} and \textbf{N}) Same as (K and L) but for normal emission [1$\bar{1}$0] photoelectron spins under 55~eV photons.}
    \label{fig:fig_geometry_B_GZ_G_S1-10}
\end{figure}

Finally, we experimentally investigate the photoelectron spin polarization along the third axis, in-plane [1$\bar{1}$0]. To this end, we select the measurement Geometry B, where even the extrinsic effects associated with the ARPES selection rules cannot cause a non-zero net spin polarization $P_{[1\bar{1}0]}$ in the paramagnetic phase. Similar to the experiments for main text Fig.~\ref{fig:fig4}, the 2~nm epitaxial RuO$_2$/TiO$_2$ film was magnetized by an in-plane magnetic field of 0.2~T pointing 45$^\circ$ between the [001] and the [1$\bar{1}$0] film axes before the spin-resolved ARPES experiments. The measured Fermi surface and $\bar{\Gamma}-\bar{Z}$ electronic band structure using 62~eV $p$-polarized photons are shown in Fig.~\ref{fig:fig_geometry_B_GZ_G_S1-10} (A and B), where the three momenta along $\bar{\Gamma}-\bar{Z}$ for spin-resolved experiments are marked by magenta and cross symbols. First under 62~eV photons, the spin-resolved EDCs at the three respective momenta are shown in Fig.~\ref{fig:fig_geometry_B_GZ_G_S1-10} (C, D, K). At finite momenta, the converted $P_{[1\bar{1}0]}$ exhibits a dominant (001)-mirror-odd behavior in Fig.~\ref{fig:fig_geometry_B_GZ_G_S1-10} (E and F). However, an even component also shows up near the Fermi level. This can be also be seen in Fig.~\ref{fig:fig_geometry_B_GZ_G_S1-10} (G to J) for the same measurements employing 55~eV photons. At normal emission, both 62~eV and 55~eV data in Fig.~\ref{fig:fig_geometry_B_GZ_G_S1-10} (K to N) show a small but finite $P_{[1\bar{1}0]}$, almost on par with the size of the experimental error bars. 

\subsubsection*{Group theory analysis}

Bulk single crystal RuO$_2$ has space group $P4_2/mnm$ (\#136). On a single Ru ion, the local site symmetry is $mmm$ ($D_{2h}$), with generators: $m_{001}$, $m_{1\bar{1}0}$, $m_{110}$. The space group further includes the non-symmorphic operations connecting the two Ru ions: $\{4_{001}|(\frac{1}{2},\frac{1}{2},\frac{1}{2})\}$ (screw rotation $4_2$) and $\{m_{100}|(\frac{1}{2},\frac{1}{2},\frac{1}{2})\}$ (glide mirror $n$). The condensation of dipoles along the $[001]$ direction ($B^-_{1g}$ irrep of $mmm$) with compensated magnetic order ($m\Gamma_2^+$ irrep of $P4_2/mnm.1^\prime$) yields the magnetic space group $P4_2^\prime/mnm^\prime$. In this group $4_2$, $m_{1\bar{1}0}$ and $m_{110}$ must be followed by time-reversal (TR) symmetry. By enforcing the magnetic point group symmetries, we find the allowed spin-splittings for this group (up to quadratic order in $\mathbf{k}$) to be
\begin{equation}
\label{eq:c_4mmm}
    k_xk_y\sigma_z,
\end{equation}
\begin{equation}
\label{eq:nc_4mmm}
    k_yk_z\sigma_x+k_xk_z\sigma_y,
\end{equation}
where we used coordinates $x$, $y$, $z$ aligned with the axes $[100]$, $[010]$ and $[001]$. These are the well known altermagnetic $d$-wave spin-splittings. The plus sign in Eq.~\ref{eq:nc_4mmm} refers to the fact that the splittings $k_yk_z\sigma_x$ and $k_xk_z\sigma_y$ must have the same amplitude, whereas $k_xk_y\sigma_z$ is free to have a different amplitude.

The epitaxial strain $\varepsilon_{xy}$ causes the symmetry to lower to $Cmmm$ (\#65), where the non-symmorphic operations $4_2$ and $n$ are now forbidden. Moreover, for very thin films, the system becomes polar due to the breaking of the mirror $m_{110}$ into $2_{110}$. Thus the space group symmetry gets further lowered to $Amm2$ (\#38) with point group $mm2$ ($C_{2v}$). For simplicity of notation, we now rotate into a coordinate system with axes aligned with the conventional cell: $[1\bar{1}0]$ as the new $x$ axis, $[001]$ as the new $y$ axis and $[110]$ as the new $z$ axis.

\begin{figure}
    \centering
    \includegraphics[width=0.6\linewidth]{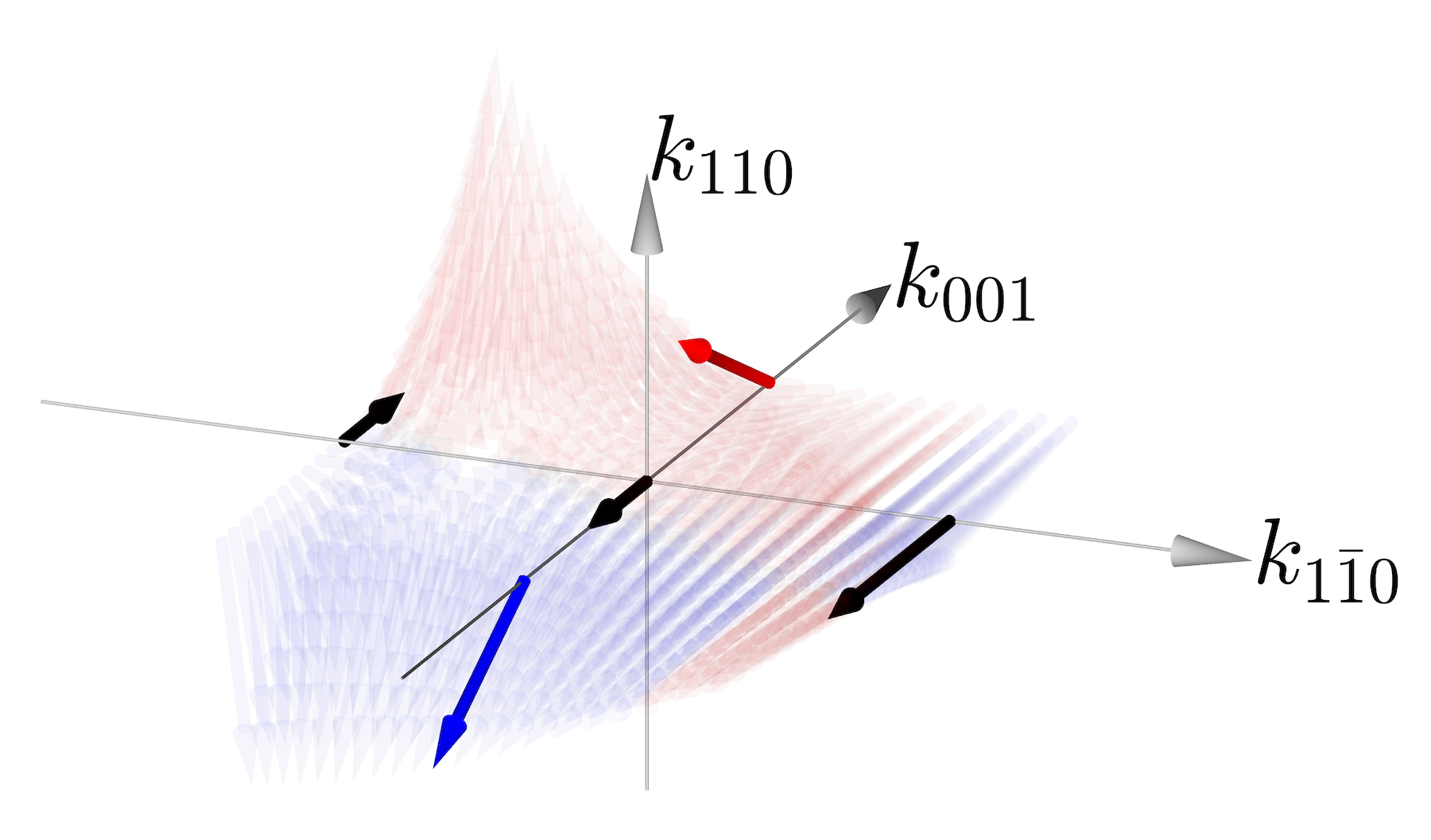}
    \caption{\textbf{Three-dimensional low-energy spin texture derived from the experiment-informed group theory analysis up to order 2 in $\mathbf{k}$.}}
    \label{fig:fig_S_theory}
\end{figure}

We assume that we can describe the magnetic phase of thin-film RuO$_2$ by starting from the paramagnetic group $mm2.1^\prime$. Because we do not know \textit{a priori} what the primary magnetic order parameter is, we list all of the possible symmetry breaking terms and label them by an irrep of $mm2.1^\prime$. In Table.~\ref{tab:mm2_table} we show the character table of the point group $mm2$, where for brevity we do not separate the TR odd and TR even irreps. By taking the products of the irreps, we construct the spin splitting terms up to quadratic order in $\mathbf{k}$, which are shown in Table.~\ref{tab:irreps} of the main text. The resultant spin-texture of a possible experiment-informed magnetic point group $m'm2'$ is visualized in Fig.~\ref{fig:fig_S_theory}.

\newpage

\clearpage

\clearpage

\clearpage

\clearpage

\clearpage

\clearpage

\begin{table}
\centering
\begin{tabular}{c|ccc|c|c|c}
  $mm2.1^\prime$       & $2_{110}$ & $m_{1\bar{1}0}$ & $m_{001}$ & $\sigma_{i}$ & $k_{i}$ & $k_ik_j$   \\ \hline
$A_1^\pm$ & \;1 & \;1 & \;1 & $\cdot$ & $k_{110}$   & $k_{1\bar{1}0}^2-k_{001}^2,\;k_{110}^2$ \\

$A_2^\pm$ & \;1 & -1  & -1  & $\sigma_{110}$   & $\cdot$ & $k_{1\bar{1}0}\,k_{001}$ \\

$B_1^\pm$ & -1  & -1 & \;1  & $\sigma_{001}$   & $k_{1\bar{1}0}$   & $k_{110}\,k_{1\bar{1}0}$ \\

$B_2^\pm$ & -1  & \;1  & -1 & $\sigma_{1\bar{1}0}$   & $k_{001}$   & $k_{110}\,k_{001}$ 
\end{tabular}
\caption{\textbf{Character table for the magnetic point group $mm2.1'$ the point group of strained, polar RuO$_2$.} Columns indicate symmetry eigenvalues under each generator, and allowed linear/quadratic $k$- and $\sigma$-terms. The superscript $\pm$ denotes whether the irrep is even/odd under time reversal $1'$. Note that the Geometry A used in our experiments breaks the $m_{001}$ mirror, and as such generates terms that transform as $B_2^+$. Similarly, Geometry B breaks the $m_{1\bar{1}0}$ mirror, generating terms that transform as $B_1^+$. }
\label{tab:mm2_table}
\end{table}
\clearpage



\noindent\textbf{File S1} DFT (PBE) relaxed crystal structure of the 15-layer inversion-symmetric slab of RuO$_2$.

\noindent\textbf{File S2} DFT (PBE) relaxed crystal structure of the 29-layer inversion symmetric slab of the RuO$_2$/TiO$_2$ heterostructure.

\clearpage 





\end{document}